\documentclass[twoside,twocolumn,9pt]{article}
\usepackage{extsizes}
\usepackage[super,sort&compress,comma]{natbib} 
\usepackage[version=3]{mhchem}
\usepackage[left=1.5cm, right=1.5cm, top=1.785cm, bottom=2.0cm]{geometry}
\usepackage{balance}
\usepackage{mathptmx}
\usepackage{sectsty}
\usepackage{graphicx} 
\usepackage{lastpage}
\usepackage[format=plain,justification=justified,singlelinecheck=false,font={stretch=1.125,small,sf},labelfont=bf,labelsep=space]{caption}
\usepackage{float}
\usepackage{fancyhdr}
\usepackage{fnpos}
\usepackage[english]{babel}
\addto{\captionsenglish}{%
  
}
\usepackage{array}
\usepackage{droidsans}
\usepackage{charter}
\usepackage[T1]{fontenc}
\usepackage[usenames,dvipsnames]{xcolor}
\usepackage{setspace}
\usepackage[compact]{titlesec}
\usepackage{hyperref}

\usepackage{cleveref}
\usepackage{subcaption}
\usepackage{appendix}

\usepackage{placeins}
\usepackage{amssymb}

\usepackage{tikz}
\usetikzlibrary{calc,positioning,arrows.meta}

\definecolor{cream}{RGB}{222,217,201}

\newcommand\norm[1]{\left\lVert#1\right\rVert}

\def \case#1#2#3{case A{#1}-T{#3}-D{#2}}
\def \casenl#1#2#3{A{#1}-T{#3}-D{#2}}


\begin{document}

\pagestyle{fancy}
\thispagestyle{plain}
\fancypagestyle{plain}{
\renewcommand{\headrulewidth}{0pt}
}

\makeFNbottom
\makeatletter
\renewcommand\LARGE{\@setfontsize\LARGE{15pt}{17}}
\renewcommand\Large{\@setfontsize\Large{12pt}{14}}
\renewcommand\large{\@setfontsize\large{10pt}{12}}
\renewcommand\footnotesize{\@setfontsize\footnotesize{7pt}{10}}
\makeatother

\renewcommand{\thefootnote}{\fnsymbol{footnote}}
\renewcommand\footnoterule{\vspace*{1pt}%
\color{cream}\hrule width 3.5in height 0.4pt \color{black}\vspace*{5pt}} 
\setcounter{secnumdepth}{5}

\makeatletter 
\renewcommand\@biblabel[1]{#1}            
\renewcommand\@makefntext[1]%
{\noindent\makebox[0pt][r]{\@thefnmark\,}#1}
\makeatother 
\renewcommand{\figurename}{\small{Fig.}~}
\sectionfont{\sffamily\Large}
\subsectionfont{\normalsize}
\subsubsectionfont{\bf}
\setstretch{1.125} 
\setlength{\skip\footins}{0.8cm}
\setlength{\footnotesep}{0.25cm}
\setlength{\jot}{10pt}
\titlespacing*{\section}{0pt}{4pt}{4pt}
\titlespacing*{\subsection}{0pt}{15pt}{1pt}

\fancyfoot{}
\fancyfoot[LO,RE]{\vspace{-7.1pt}\includegraphics[height=9pt]{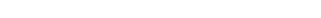}}
\fancyfoot[CO]{\vspace{-7.1pt}\hspace{13.2cm}\includegraphics{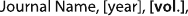}}
\fancyfoot[CE]{\vspace{-7.2pt}\hspace{-14.2cm}\includegraphics{head_foot/RF}}
\fancyfoot[RO]{\footnotesize{\sffamily{1--\pageref{LastPage} ~\textbar  \hspace{2pt}\thepage}}}
\fancyfoot[LE]{\footnotesize{\sffamily{\thepage~\textbar\hspace{3.45cm} 1--\pageref{LastPage}}}}
\fancyhead{}
\renewcommand{\headrulewidth}{0pt} 
\renewcommand{\footrulewidth}{0pt}
\setlength{\arrayrulewidth}{1pt}
\setlength{\columnsep}{6.5mm}
\setlength\bibsep{1pt}

\makeatletter 
\newlength{\figrulesep} 
\setlength{\figrulesep}{0.5\textfloatsep} 

\newcommand{\topfigrule}{\vspace*{-1pt}%
\noindent{\color{cream}\rule[-\figrulesep]{\columnwidth}{1.5pt}} }

\newcommand{\botfigrule}{\vspace*{-2pt}%
\noindent{\color{cream}\rule[\figrulesep]{\columnwidth}{1.5pt}} }

\newcommand{\dblfigrule}{\vspace*{-1pt}%
\noindent{\color{cream}\rule[-\figrulesep]{\textwidth}{1.5pt}} }

\makeatother

\twocolumn[
  \begin{@twocolumnfalse}
{\includegraphics[height=30pt]{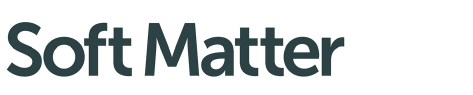}\hfill\raisebox{0pt}[0pt][0pt]{\includegraphics[height=55pt]{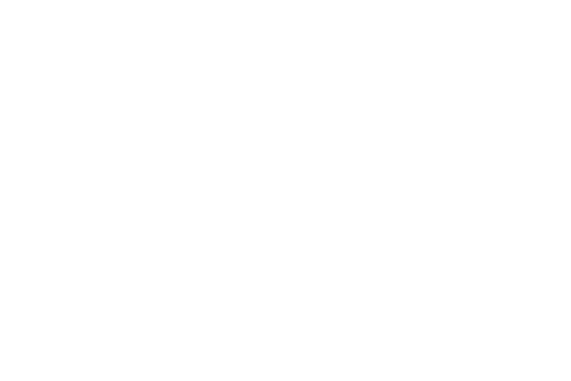}}\\[1ex]
\includegraphics[width=18.5cm]{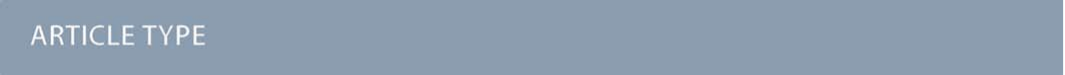}}\par
\vspace{1em}
\sffamily
\begin{tabular}{m{4.5cm} p{13.5cm} }

\includegraphics{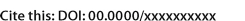} & \noindent\LARGE{\textbf{Benchmarking the flow of epithelial cell monolayer with self-aligning deformable active membranes   $^\dag$}} \\
\vspace{0.3cm} & \vspace{0.3cm} \\

 & \noindent\large{Marcos Pasa\textit{$^{a}$}, {Carine P. Beatrici}\textit{$^{a}$}, {François Graner}\textit{$^{c}$}, Leonardo G. Brunnet\textit{$^{a}$} and {Emanuel F. Teixeira}\textit{$^{b}$}} \\

\includegraphics{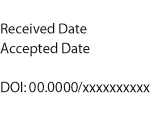} & \noindent\normalsize{Collective cell migration emerges from the interplay between motility, deformability and mechanical interactions, yet incorporating these ingredients into computationally efficient tissue models remains challenging. Here, we benchmark self-aligning deformable active membranes in a confined-flow geometry that mimics epithelial monolayer migration around a circular obstacle. In this model, cells are represented as deformable, adhesive membranes whose self-propulsion direction relaxes toward their velocity. By systematically varying the self-alignment timescale, cell–cell adhesion and inlet forcing, we characterize the resulting flows through collective alignment, relative density, neighbor rearrangements and spatial velocity fields. The model captures a broad spectrum of tissue behaviors, ranging from disordered, liquid-like flows to highly aligned, solid-like states. Compared with a related multiparticle model, self-aligning active membranes achieve stronger collective alignment, exhibit a more systematic density response and access states closer to both limits of the solid–liquid spectrum. We further show that increasing the target shape index promotes cell elongation and accelerates tissue flow, directly linking cell-scale deformability to tissue-scale transport. Finally, we compare simulated velocity profiles with experimental measurements from \textit{in vitro} migrating MDCK epithelial cell monolayers and find qualitative agreement across multiple horizontal and vertical transects around the obstacle. These results establish self-aligning deformable active membranes as a versatile framework for connecting cell mechanics, shape adaptation and self-alignment to collective tissue migration in confined geometries.
} \\

\end{tabular}

 \end{@twocolumnfalse} \vspace{0.6cm}

  ]

\renewcommand*\rmdefault{bch}\normalfont\upshape
\rmfamily
\section*{}
\vspace{-1cm}


\footnotetext{\textit{$^{a}$~Instituto de Física, Universidade Federal do Rio Grande do Sul}}
\footnotetext{\textit{$^{b}$~Huygens-Kamerlingh Onnes Laboratory, Universiteit Leiden, PO Box 9504, 2300 RA Leiden, the Netherlands.}}
\footnotetext{\textit{$^{c}$~Université Paris Cité, CNRS, Matière et Systèmes Complexes, F-75013 Paris, France.}}

\section{Introduction}
Tissue mechanical properties play a fundamental role in embryonic development \cite{huang_exploring_2024,petridou_tissue_2019}, wound healing \cite{ding_microfluidic-based_2025}, and cancer progression \cite{massey_mechanical_2024,bera_extracellular_2022}. Mechanical biomarkers have also emerged as tools for early disease detection, as mechanical alterations can precede changes in biochemical markers \cite{eroles_advances_2023}. The mechanical response of individual cells is largely determined by the cytoskeleton, plasma membrane, and nucleus \cite{galie_how_2022}, while tissue-level behavior additionally depends on intercellular adhesion and interactions with the extracellular environment \cite{schiele_actin_2015}. Tissue morphology and dynamics thus emerge from intracellular mechanics and collective cell-cell interactions.
\begin{figure*}[!t]
    \centering

    \begin{minipage}[t]{0.68\textwidth}
        \centering

 \begin{subfigure}[t]{\linewidth}
    \centering
    \includegraphics[
        width=0.93\linewidth,
    ]{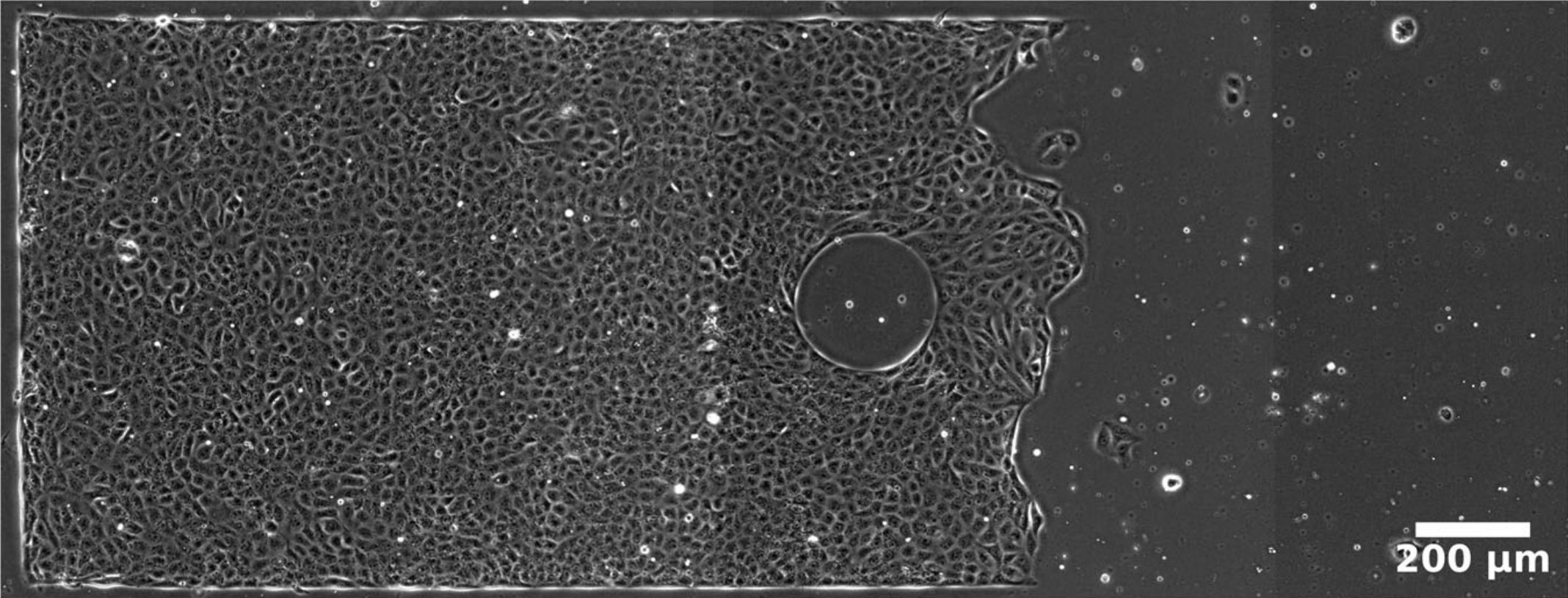}
    \caption{Experiment}
    \label{fig:melina}
\end{subfigure}
        \vspace{0.8em}

        \begin{subfigure}[t]{\linewidth}
            \centering
            \begin{minipage}[c][0.30\textwidth][c]{\linewidth}
                \centering
                \makebox[\linewidth][c]{%
                    \includegraphics[
                        width=1.08\linewidth,
                        height=0.30\textwidth,
                        keepaspectratio
                    ]{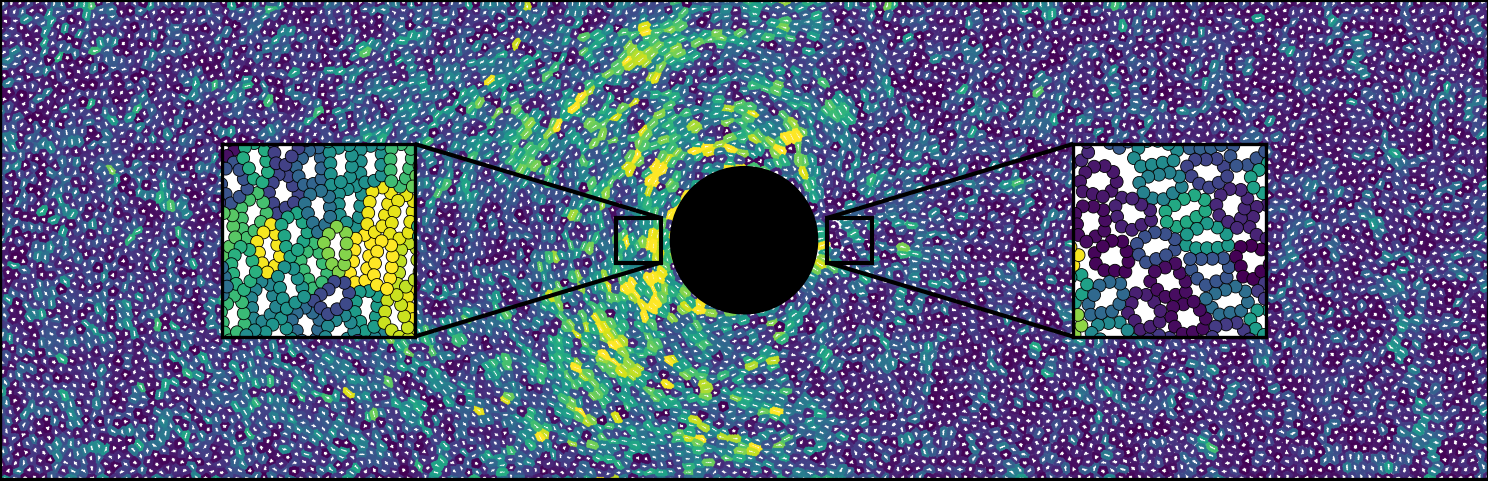}
                }
            \end{minipage}
            \caption{Simulation}
            \label{fig:channel_view_rings}
        \end{subfigure}
    \end{minipage}
    \hfill
    \begin{minipage}[t]{0.29\textwidth}
        \centering

        \begin{subfigure}[t]{\linewidth}
            \centering
            \includegraphics[
                width=\linewidth
            ]{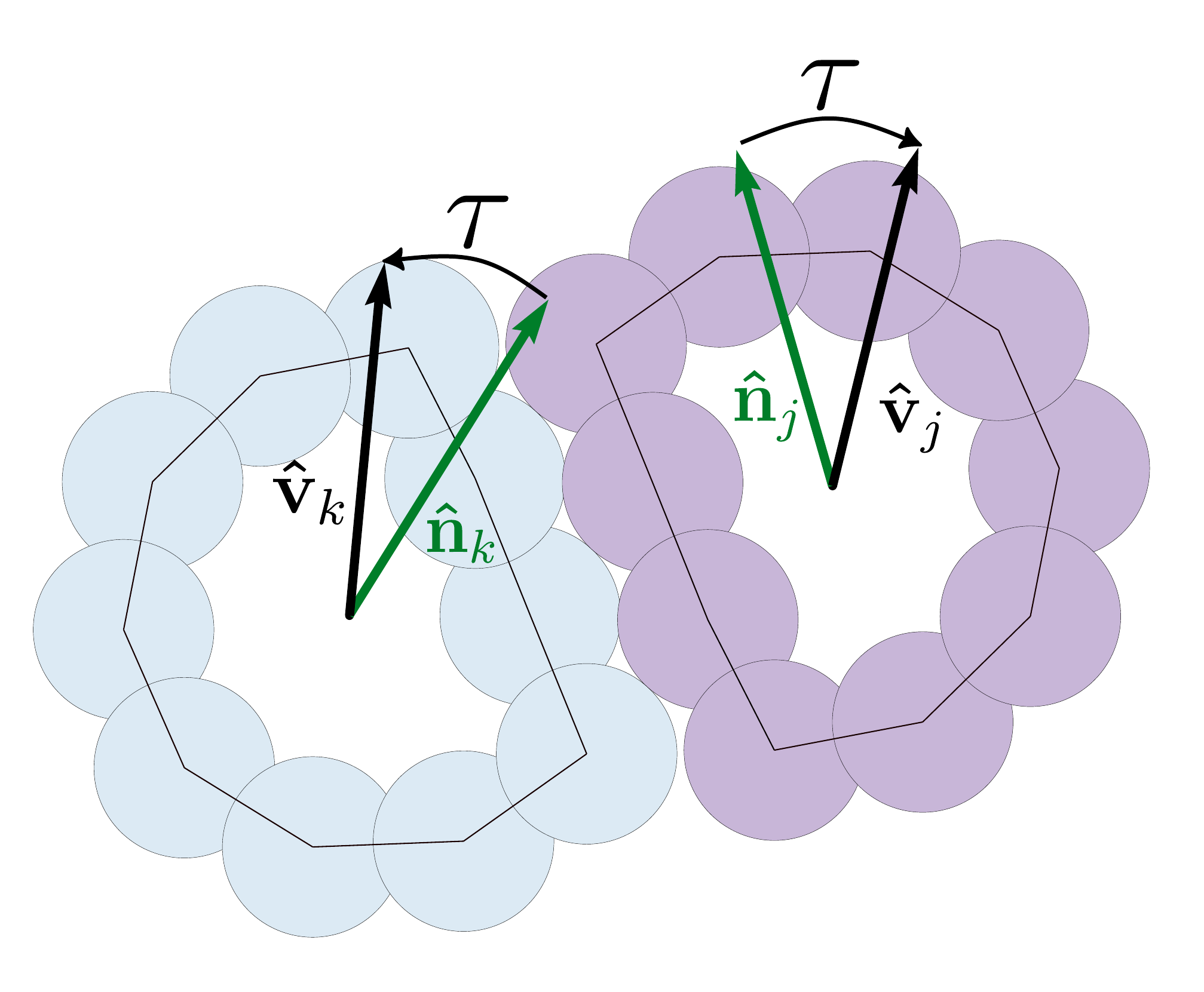}
            \caption{Self-alignment}
            \label{fig:model_pol_scheme}
        \end{subfigure}

        \vspace{0.8em}

        \begin{subfigure}[t]{\linewidth}
            \centering
            \includegraphics[
                width=\linewidth
            ]{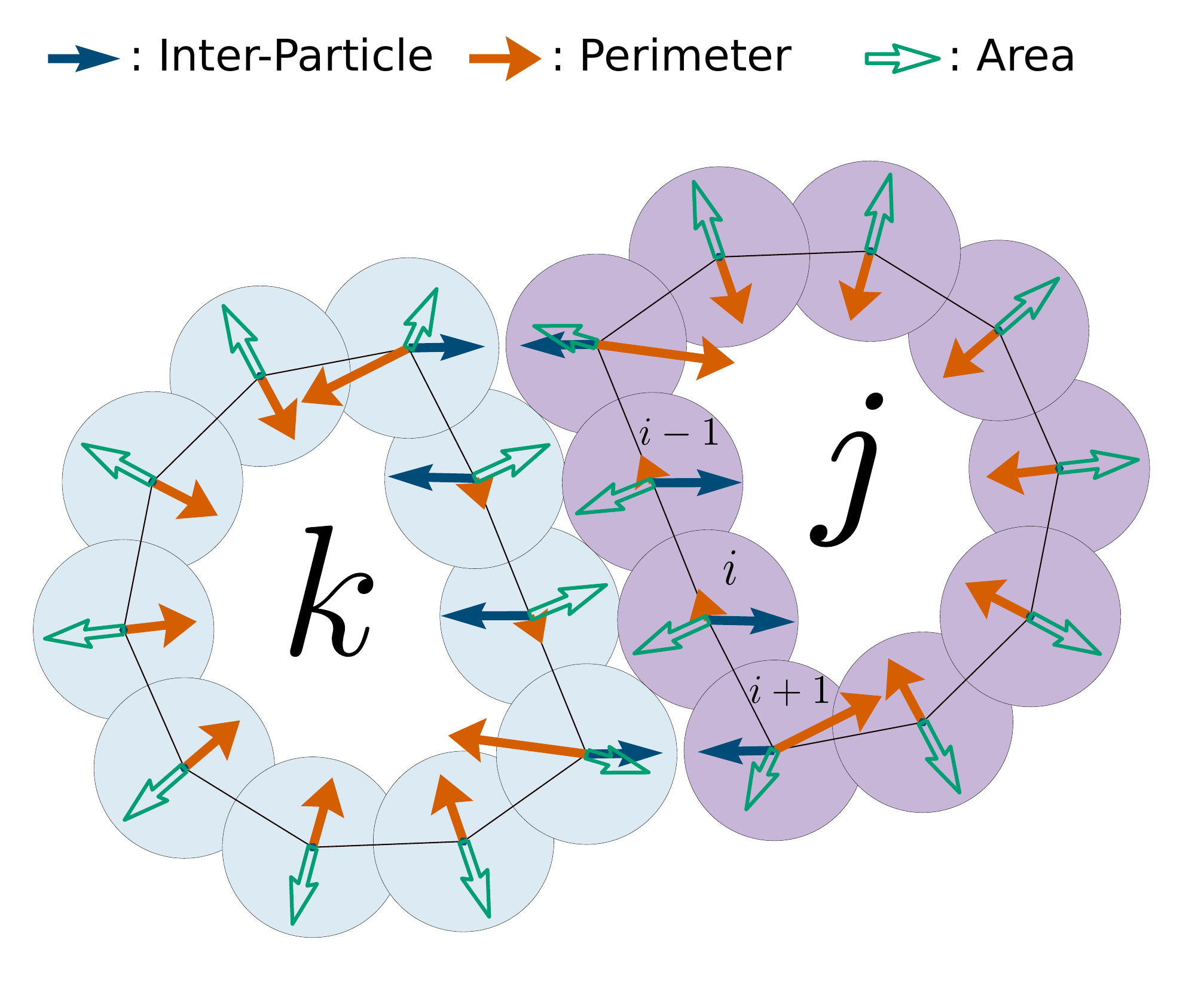}
            \caption{Forces}
            \label{fig:model_scheme}
        \end{subfigure}
    \end{minipage}

    \caption{
        Model used to simulate epithelial monolayer migration in Stokes flow.
        Panel~(\subref{fig:melina}) shows a snapshot acquired during the
        experimental procedure of Duprez and colleagues~\cite{duprez2026geometric}. Panel~(\subref{fig:channel_view_rings}) shows a channel filled with active membranes. Membranes are colored by their asphericity, a scalar measure of shape anisotropy ranging from 0 for more circular membranes (dark purple) to 1 for more elongated membranes (yellow). Insets show magnified views of the channel region immediately before and after the obstacle. Panels~(\subref{fig:model_pol_scheme}) and
        (\subref{fig:model_scheme}) illustrate two interacting active membranes,
        labeled $j$ and $k$. Each membrane is depicted as a sequence of particles
        of diameter $\sigma$ [\cref{eq:energy_c_adh}], shown as circles of the
        same color, with black lines representing the springs connecting
        neighboring particles.
        Panel~(\subref{fig:model_pol_scheme}) illustrates the self-alignment
        mechanism: the membrane polarization $\mathbf{\hat n}$ relaxes toward the
        direction of the membrane velocity $\mathbf{\hat v}$ over the
        characteristic timescale $\tau$ [\cref{eq:eq_motion}].
        Panel~(\subref{fig:model_scheme}) illustrates the forces acting on the particles. Orange arrows represent spring forces [\cref{eq:energy_p}], green arrows represent forces arising from the preferred-area energy [\cref{eq:energy_a}], and blue arrows represent interparticle interactions [\cref{eq:energy_c_adh}].
    }
    \label{fig:fig1}
\end{figure*}
A useful geometry for probing epithelial tissue mechanics consists of a confluent cell monolayer migrating along an adhesive strip containing a central non-adhesive circular obstacle~\cite{tlili_migrating_2020,chhabra_stokes_2024,duran_2020,duprez2026geometric}. Migration is initiated by creating a free front, which drives the tissue into the initially unoccupied region and around the obstacle. Inspired by Stokes's classical problem of viscous flow past an obstacle~\cite{stokes_effect_1851}, this configuration has also been used to characterize complex materials with elastic, plastic, and viscous responses, including liquid foams~\cite{cheddadi_understanding_2011}. In epithelial monolayers, the resulting velocity field provides a controlled way to probe how geometric confinement and boundary conditions shape collective migration and the tissue-scale mechanical response~\cite{duprez2026geometric}.

Collective cell migration is commonly described using active matter models ~\cite{giavazzi2018flocking,kammeraat2025correlated,teixeira_collective_2026}, where cells are self-propelled entities with an internal degree of freedom interpreted as cell polarity. Collective motion can emerge from explicit alignment interactions between neighboring cells, as in the Vicsek model \cite{vicsek_novel_1995}, or from self-alignment mechanisms, where polarity aligns with velocity \cite{PhysRevLett.76.3870,PhysRevE.74.061908,PhysRevE.84.040301,baconnier_self-aligning_2025}. Unlike Vicsek-type alignment, self-alignment does not require cells to explicitly average the orientations of their neighbors \cite{PhysRevE.74.061908}. Instead, collective behavior arises from feedback between cell motion, polarity, and local mechanical interactions.

Modeling strategies for collective cell migration span multiple levels of detail. Early models represented cells as active soft disks  \cite{PhysRevE.74.061908,belmonte_self-propelled_2008,gregoire_moving_2003,teixeira_collective_2026}. To capture the geometry of confluent tissues, vertex models represent cells as polygonal domains whose vertices constitute the fundamental degrees of freedom~\cite{bi_density-independent_2015,barton_active_2017,rozman2024cell}, whereas Voronoi models define cell shapes through a tessellation generated from a set of cell-center points~\cite{bi_motility-driven_2016}. Other approaches include Cellular Potts models \cite{kabla_collective_2012,kafer_moving_2006}, phase-field descriptions~\cite{loewe_solid-liquid_2020,balasubramaniam2021investigating}, and sub-cellular elements, where each cell comprises interacting internal elements that allow explicit shape deformations \cite{newman_modeling_2005}.

Deformable particle models explicitly incorporate key cellular mechanical properties, including cell deformability, friction, cell-boundary curvature, and cell-cell interfaces~\cite{mkrtchyan2014new,madhikar2021jamming,PhysRevLett.121.248003,treado2022localized,PhysRevMaterials.5.055605,ourique_modelling_2022,D2SM01079H,PhysRevLett.130.130002,theeyancheri2024dynamic,PhysRevResearch.6.L012036,ray2025modeling,mackeith2025deformable}. Within this class, Teixeira \textit{et al.} \cite{teixeira_single_2021} introduced the Deformable Active Membrane Model (DAMM), where each cell is a membrane of self-aligning active particles connected by harmonic springs. The DAMM has since been used to study segregation driven by differences in local contractility \cite{teixeira_segregation_2025}, providing a framework for connecting microscopic mechanics to emergent tissue behavior.
A distinctive feature of the DAMM is its explicit representation of cell boundaries. Unlike Vertex and Voronoi-based models, where neighboring cells share a single geometric interface, the DAMM assigns an independent membrane to each cell, allowing interactions to occur directly between adjacent cell membranes. This provides a more detailed description of cell-cell mechanical contacts and their role in collective tissue dynamics.
\begin{figure*}[!t]
    \centering
    \includegraphics[width=\textwidth]{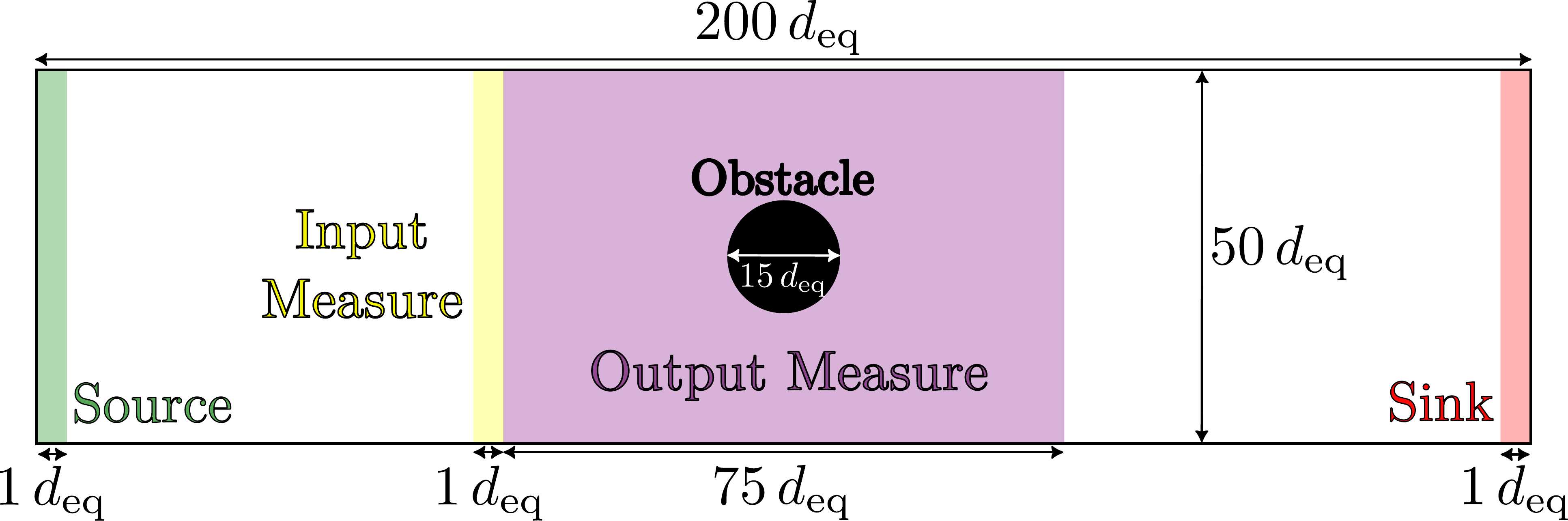}
    \caption{
    Illustration of the channel geometry, with dimensions expressed in units of the equilibrium membrane diameter $d_{eq}$ [\cref{eq:ring_diamenter_eq}]. The green region indicates where membranes are created, the red region where they are removed, the yellow region denotes the input measurement region, and the purple region denotes the output measurement region. The black circle represents the obstacle.
    }
    \label{fig:channel_view}
\end{figure*}
Beatrici \textit{et al.} \cite{beatrici_comparing_2023} systematically benchmarked five active matter models for collective cell migration in the Stokes geometry, comparing models with increasing mechanical and geometrical detail. Among these, their study included a model similar to the DAMM. However, the constituent particles follow Vicsek-like dynamics, and each cell contains a central particle representing the nucleus. To extend this comparison, we apply the DAMM to the same benchmark and analyze how deformability affects collective migration in confined flow.

We show that, under adequate parametrization, the DAMM promotes strong spatial cohesion and  robust collective ordering. By systematically varying the target shape index $p_0$, which controls  membrane deformability and shape, we demonstrate that increased deformability leads to faster flow, highlighting the central role of mechanical properties in collective migration. To validate the model, we compare our simulations with experimental measurements of epithelial cell migration in the Stokes geometry from Durande \cite{duran_2020}. Analysis of velocity field profiles shows that the DAMM reproduces key experimental observations over a broad range of parameters.  All simulations were performed using Phystem (see \Cref{app:phystem} of the Supplementary Information (SI) for more details), an open-source package combining a user-friendly Python interface with a high-performance C++ core for efficient large-scale simulations.

The paper is organized as follows. Section~2 presents the model. Section~3 introduces the Stokes geometry and flow protocol. Sections~4 and~5 describe the parameter-exploration strategy and measured observables. Sections~6--8 report the benchmark results, examine how membrane deformability affects shape and flow, and compare the simulations with experimental data. Section~9 concludes.

\section{Model} \label{sec:model}
The model employed in this work is based on the active membrane model introduced by Teixeira \textit{et al.}~\cite{teixeira_segregation_2025}. The system consists of $N$ active membranes, each composed of $n$ active particles connected by elastic bonds. We extend the original formulation by adapting the self-alignment mechanism introduced by Szabó \textit{et al.}~\cite{szabo_phase_2006} for elastic disks. In the present model, this mechanism acts on a single global polarity associated with each membrane. Unlike earlier implementations in which each constituent particle possessed an independent polarity~\cite{teixeira_single_2021,ourique_modelling_2022}, all particles within a given membrane share a common polarity, as in Refs.~\cite{teixeira_segregation_2025,gopinath2025cell}. This global polarity determines the overall self-propulsion direction of the membrane and aligns with its center-of-mass velocity rather than with the velocity of each individual particle. 
This modeling choice provides a more realistic representation of cell polarity. Accordingly, the dynamics of each particle are described in two dimensions by the following set of coupled overdamped equations:
\begin{equation}
\begin{aligned}
\mathbf{\dot{r}}_{i,j} &= v_{0}\;\mathbf{\hat{n}}_{j} +\mu\,\mathbf{F}_{i,j} \\
\dot{\theta}_{j} &= \frac{1}{\tau}\sin^{-1}\bigg(\mathbf{\hat{n}}_{j} \times \mathbf{\hat{v}}_{j} \cdot \mathbf{\hat e}_z \bigg) + \sqrt{2 D_R}\xi_{j}(t)
\end{aligned}
\label{eq:eq_motion}
\end{equation}
where $\mathbf{r}_{i,j}$ denotes the position of the $i$-th particle belonging to the $j$-th membrane at time $t$, $\mu$ is the mobility, and $v_0$ is the active speed whose orientation is set by the membrane polarity $\hat{\mathbf{n}}_j$. The second term, $\mathbf{F}_{i,j} = -\boldsymbol{\nabla}_{i,j} E$, represents the total force acting on the $i$-th particle of the $j$-th membrane, obtained from the interaction energy $E$. The polarity angle of the membrane, denoted by $\theta_j$, tends to align with the direction of the membrane center-of-mass velocity, $\mathbf{\hat v}_j = \mathbf{v}_j / |\mathbf{v}_j|$, where $\mathbf{v}_j = \langle \mathbf{\dot{r}}_{i,j} \rangle_i$, over a characteristic relaxation time $\tau$. In addition, this internal self-alignment process is subject to random fluctuations, which are described by an angular Gaussian white noise $\xi_j$  with correlation $\left \langle \xi_{j}(t_{1})\xi_{k}(t_{2}) \right \rangle = \delta_{j,k}\delta (t_{1}-t_{2})$. The noise term $D_{R}$ represents the rotational diffusion coefficient, which defines a characteristic timescale given by $\tau_R=1/D_{R}$.
The energy function has contributions from a perimeter energy ($E_P$) (springs connecting membrane neighboring particles), preferred-area energy ($E_A$), core repulsion, and intercellular adhesion among particles  ($E_{c, adh}$): 
\begin{subequations}
\label{eq:energy}
\begin{align}
E_P
&=
\frac{k_P}{2}
\sum_{j=1}^{N}
\sum_{i=1}^{n}
\left(
\left\lvert \mathbf{l}_{i,j} \right\rvert-l_0
\right)^2,
\label{eq:energy_p}
\\
E_A
&=
\frac{k_A}{2}
\sum_{j=1}^{N}
\left(A_j-A_0\right)^2,
\label{eq:energy_a}
\\
E_{c,\mathrm{adh}}
&=
\frac{k_c}{2}
\sum_{r_{ik}\leq \sigma}
\left(r_{ik}-\sigma\right)^2
+
\frac{k_{\mathrm{adh}}}{2}
\sum_{\sigma<r_{ik}\leq l_{\mathrm{adh}}}
\left(r_{ik}-\sigma\right)^2,
\label{eq:energy_c_adh}
\\
E
&=
E_P+E_A+E_{c,\mathrm{adh}}.
\label{eq:energy_total}
\end{align}
\end{subequations}
where $\mathbf{l}_{i,j} = \mathbf{r}_{i,j} - \mathbf{r}_{i-1,j}$ is the vector connecting consecutive particles in membrane $j$, $k_{P}$ is the elastic spring stiffness constant, and $l_{0}$ its target length. The elastic constant related to area control is $k_{A}$, $A_{j}$ is the inner area of the $j$-th membrane (calculated as the area of the polygon defined by the positions of its particle centers), and $A_{0}$ is the target area. The two terms in \cref{eq:energy_c_adh} represent  core repulsion and adhesion between particles from different membranes, respectively. Same-membrane particles also exhibit core repulsion beyond nearest-neighbor pairs. The parameter $k_{c}$ denotes the characteristic core repulsion interaction, $k_\mathrm{adh}$ the adhesion intensity, $\sigma$ is the equilibrium distance, which effectively defines the particle diameter, and $l_\mathrm{adh}$ is the maximum interaction distance. The quantity $r_{ik}$ is the distance between particles $i$ and $k$.  The first summation in Eq. \ref{eq:energy_c_adh} involves any pair of particles not connected by a spring, whereas the second includes only pairs belonging to different membranes. Figure \ref{fig:fig1} illustrates two interacting membranes showing the self-alignment mechanism and the forces involved.

A common limitation of the DAMM is the occurrence of inter-membrane intersections, particularly in regions of high local pressure. To resolve this, we use an additional repulsive force to undo these overlaps, as detailed in Section \ref{sec:anti_intersection} of the SI.

\section{Stokes geometry system}
\label{sec:stokes}
Following the protocol described in Beatrici \textit{et al.} \cite{beatrici_comparing_2023}, we perform simulations in a rectangular channel with a circular obstacle at its center (Stokes geometry, \cref{fig:channel_view}). Channel dimensions are expressed in terms of the equilibrium membrane diameter $d_{eq}$, which is defined as
\begin{equation} \label{eq:ring_diamenter_eq}
d_{eq} = \sqrt{A_0} + \sigma.
\end{equation}
For the parameters used in this work (\cref{tab:fix_pars}), $d_{eq} \approxeq 3.23\sigma$. Flux is induced in this geometry by introducing membranes at the inlet of the channel whenever space is available and removing them at the outlet.

While the membrane center lies within the creation region (see \cref{fig:channel_view}), a constant force of magnitude $F_i$, directed to the right, is applied to all particles composing the membrane.

\subsection{Boundary conditions in the transverse direction and at the obstacle}
The system is confined in the $y$ direction using frictionless boundary conditions, implemented by removing the outward velocity component perpendicular to the boundary.
The central obstacle is modeled through a radial elastic repulsion. Let $\mathbf{R}_{\mathrm{obs}}$ and $R$ denote the position of the obstacle center and its radius, respectively. For particle $i$ belonging to membrane $j$, we define the distance from the obstacle as $|\mathbf{R}_{i,j}| = |\mathbf{r}_{i,j}-\mathbf{R}_{\mathrm{obs}}|$. The force exerted by the obstacle on the particle is then given by \begin{equation} \mathbf{F}_{\mathrm{obs},i}^{\,j} = -k_{\mathrm{obs}} \left[ |\mathbf{R}_{i,j}| - R \right] \Theta\left( R - |\mathbf{R}_{i,j}| \right) \hat{\mathbf{R}}_{i,j}, \label{eq:obs_force} \end{equation}
where $\hat{\mathbf{R}}_{i,j} = \frac{\mathbf{R}_{i,j}}{|\mathbf{R}_{i,j}|}$, $k_{\mathrm{obs}}$ is the strength of the elastic repulsion and $\Theta$ is the Heaviside step function. The interaction occurs only when the particle overlaps with the obstacle, that is, when $|\mathbf{R}_{i,j}|<R$. The unit vector $\hat{\mathbf{R}}_{i,j}$ points radially outward from the obstacle center, ensuring that the resulting force pushes overlapping particles away from the obstacle.

\subsection{Fixed model parameters}
Throughout this work, lengths are measured in units of the particle diameter, $\sigma$, times in units of the active time scale, $\tau_0=\sigma/v_0$, and energies in units of $\epsilon=v_0\sigma/\mu$. We use $\sigma=1$, $\mu=1$, $v_0=0.5$. We express the membrane-polarity persistence time, $\tau_R=1/D_{R}$, in terms of the rotational Péclet number~\cite{martin2018collective,zhao2021phases},
\begin{equation}
\mathrm{Pe}\equiv \frac{v_0\tau_R}{\sigma}
=\frac{\tau_R}{\tau_0}.
\end{equation}
Thus, varying $\mathrm{Pe}$ is equivalent to tuning the rotational noise level of the system: smaller values of $\mathrm{Pe}$ correspond to faster polarity decorrelation, whereas larger values correspond to more persistent active motion. In this work, we fix the polarity persistence time at $\tau_R=2\tau_0$, corresponding to $\mathrm{Pe}=2$.

Each membrane is composed of $n=10$ particles connected by harmonic springs with spring constant $k_P = 56\epsilon/\sigma^2$ and equilibrium length $l_0=0.8\sigma$. The membrane area is controlled by the area-preservation constant $k_A=26\epsilon/\sigma^{4}$, with equilibrium area $A_0=4.92\sigma^2$. From these parameters, we define the dimensionless target shape index \cite{bi_density-independent_2015},
\begin{equation}
p_0=\frac{P_0}{\sqrt{A_0}},
\end{equation}
where $P_0=n l_0$ is the target perimeter of the membrane imposed by the springs. In the membrane model, there is a competition between two perimeters: one is $P_0$, and the other is the perimeter of a regular polygon with area $A_0$ and $n$ sides, which we call $P_A$, since the preferred-area energy tends to transform the membrane into a regular polygon. These two perimeters are equal when $p_0 \approx 3.6$ (for $n=10$). If $P_A > P_0$ (i.e., $p_0 \lesssim 3.6$), the membrane is less deformable and more rounded; on the other hand, if $P_A < P_0$ (i.e., $p_0 \gtrsim 3.6$), the membrane is very flexible, meaning that changes in its shape do not cost much energy. For the parameters used here, $p_0 \simeq 3.6$. In \cref{sec:po_exploration}, we explicitly vary $p_0$ to investigate the effect of membrane shape on tissue flow.

Particle-particle steric repulsion is set by $k_c=56 \epsilon/\sigma^2$, while adhesive interactions act over a distance $l_{\rm adh}=1.18\sigma$. The central obstacle force constant is $k_{\rm obs}=60\epsilon/\sigma^{2}$. 
The equations of motion are integrated using the Euler-Maruyama algorithm~\cite{kloeden1992numerical} with time step $\Delta t=0.005\tau_0$. The chosen parameter values are summarized in \cref{tab:fix_pars}.
\begin{table}[h]
    \centering
    \caption{Fixed parameters used in all simulations. Length, time, and energy are measured in units of $\sigma$, $\tau_0=\sigma/v_0$, and $\epsilon = v_0\sigma/\mu$, respectively. Parameters marked as “Ref. unit” define the system of units employed throughout this work}
    \label{tab:fix_pars}
    \begin{tabular*}{0.48\textwidth}{@{\extracolsep{\fill}}llll}
        \hline
        Parameter & Value & Unit & Description \\
        \hline 
        $\mu$ & 1 &Ref. unit & Mobility \\
        $\sigma$ & 1 & Ref. unit & Particle diameter \\
        $v_0$ & 0.5 & Ref. unit & Activity \\
        $k_P$ & $56$ & $\epsilon/\sigma^2$ & Spring constant \\
        $l_0$ & $0.8$ & $\sigma$ & Spring equilibrium length \\
        $k_c$ & $56$ & $\epsilon/\sigma^2$ & Core particle force constant \\
        $l_\mathrm{adh}$ & $1.18$ & $\sigma$ & Max. interaction distance\\
        $k_A$ & $26$ & $\epsilon / \sigma^4$ & Preferred-area energy constant \\
        $A_0$ & $4.92$ & $\sigma^2$ & Membrane equilibrium area \\
        $k_{obs}$ & $60$ & $\epsilon / \sigma^2$ & Obstacle force constant \\
        $D_R$ & $0.5$ & $1/\tau_0$ & Rotational noise \\
        $\Delta t$ & $0.005$ & $\tau_0$ & Integration time step \\
        $n$ & $10$ & - & \# of particles per membrane \\
        \hline
    \end{tabular*}
\end{table}

\begin{figure*}[h!]
    \centering

    \begin{tikzpicture}
        \node[inner sep=0] (figures) {%
            \begin{minipage}{\textwidth}
                \centering

                \begin{subfigure}{0.47\textwidth}
                    \centering
                    \includegraphics[width=\textwidth]
                    {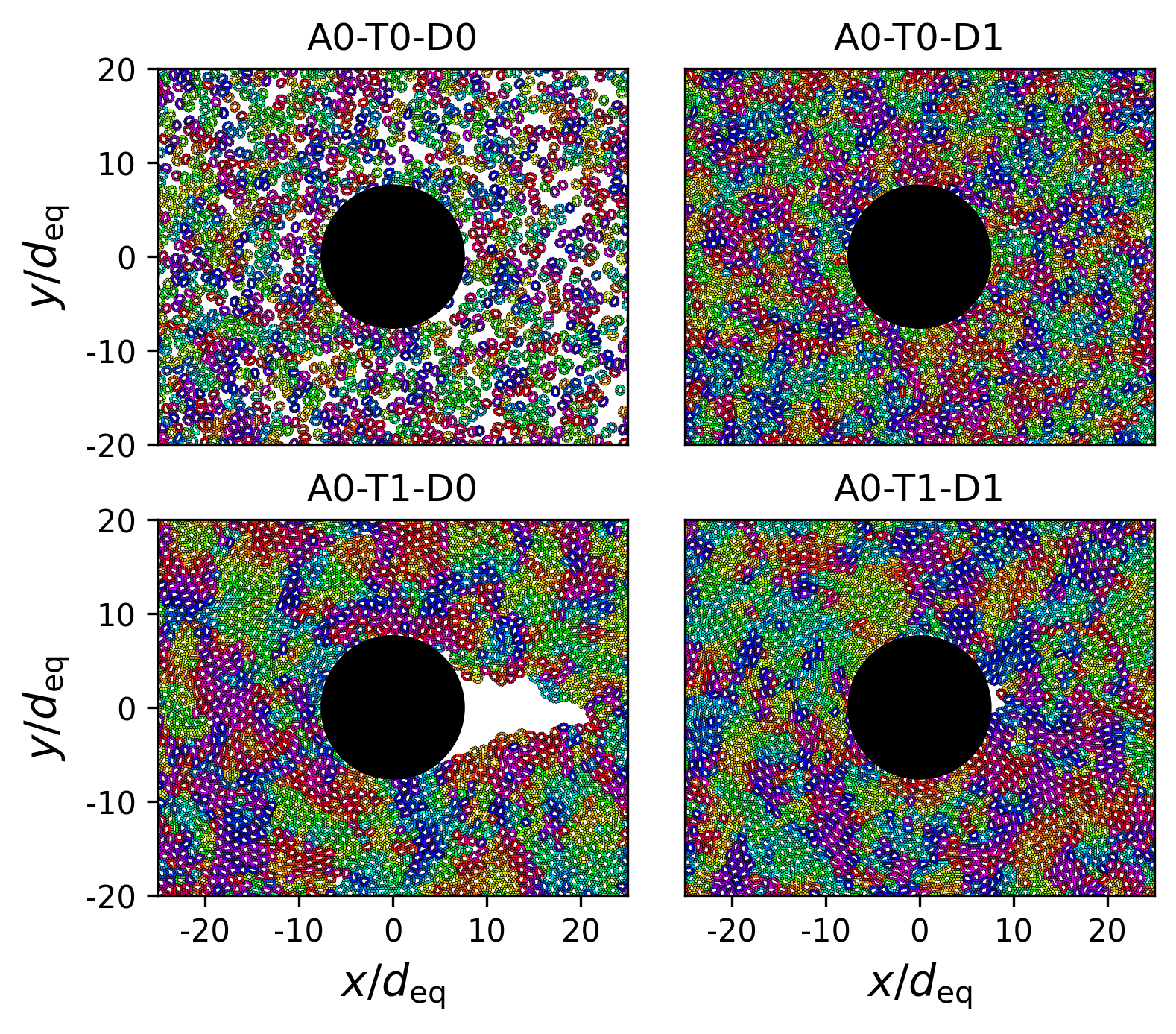}
                    \caption{High $\tau$}
                    \label{fig:low_align_snaps}
                \end{subfigure}
                \hfill
                \begin{subfigure}{0.47\textwidth}
                    \centering
                    \includegraphics[width=\textwidth]
                    {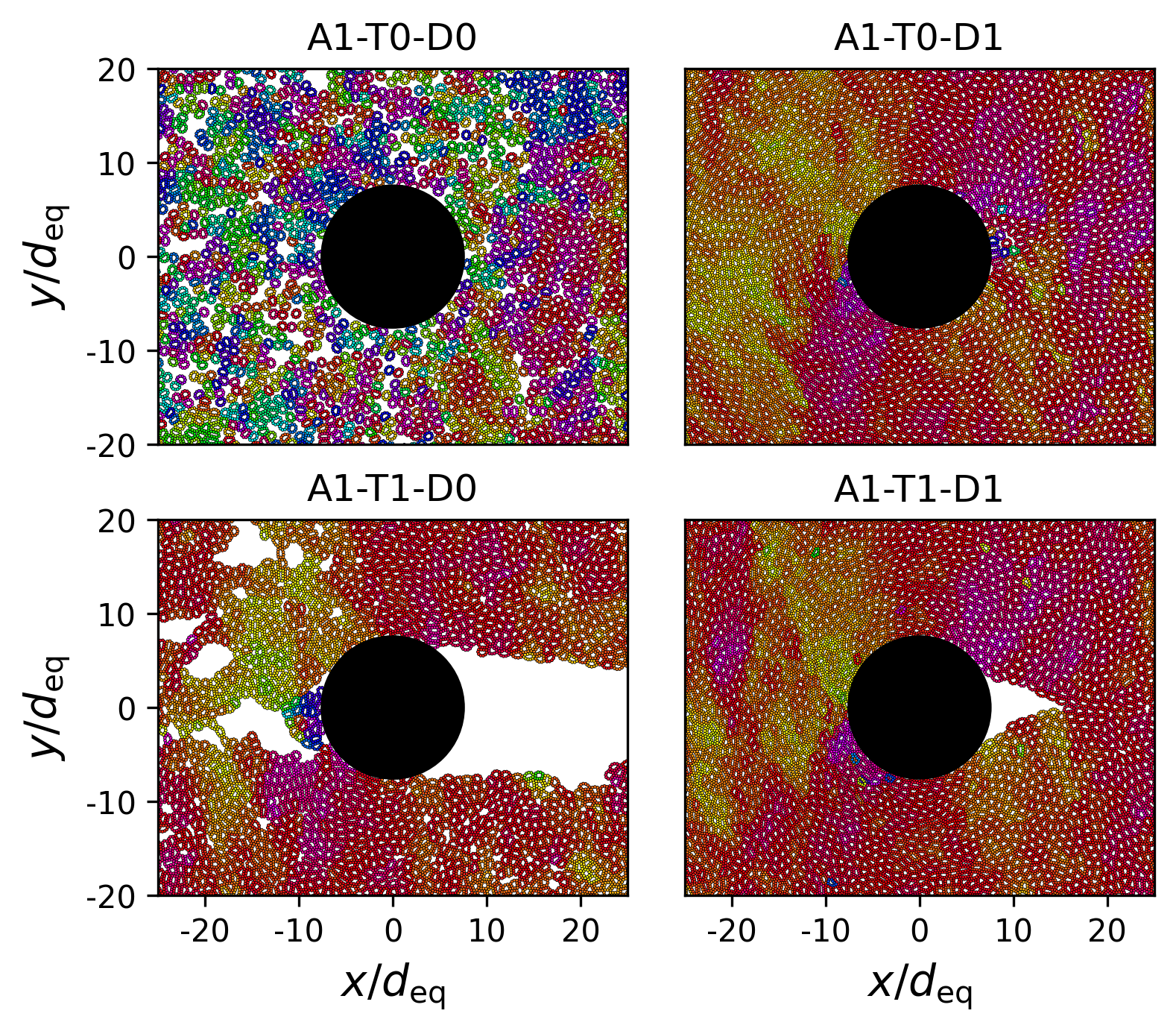}
                    \caption{Low $\tau$}
                    \label{fig:high_align_snaps}
                \end{subfigure}
            \end{minipage}%
        };

        \node[
            anchor=south,
            inner sep=0,
            yshift=1.2cm,
            xshift=0.3cm
        ] at (figures.south) {%
            \includegraphics[width=0.05\textwidth]
            {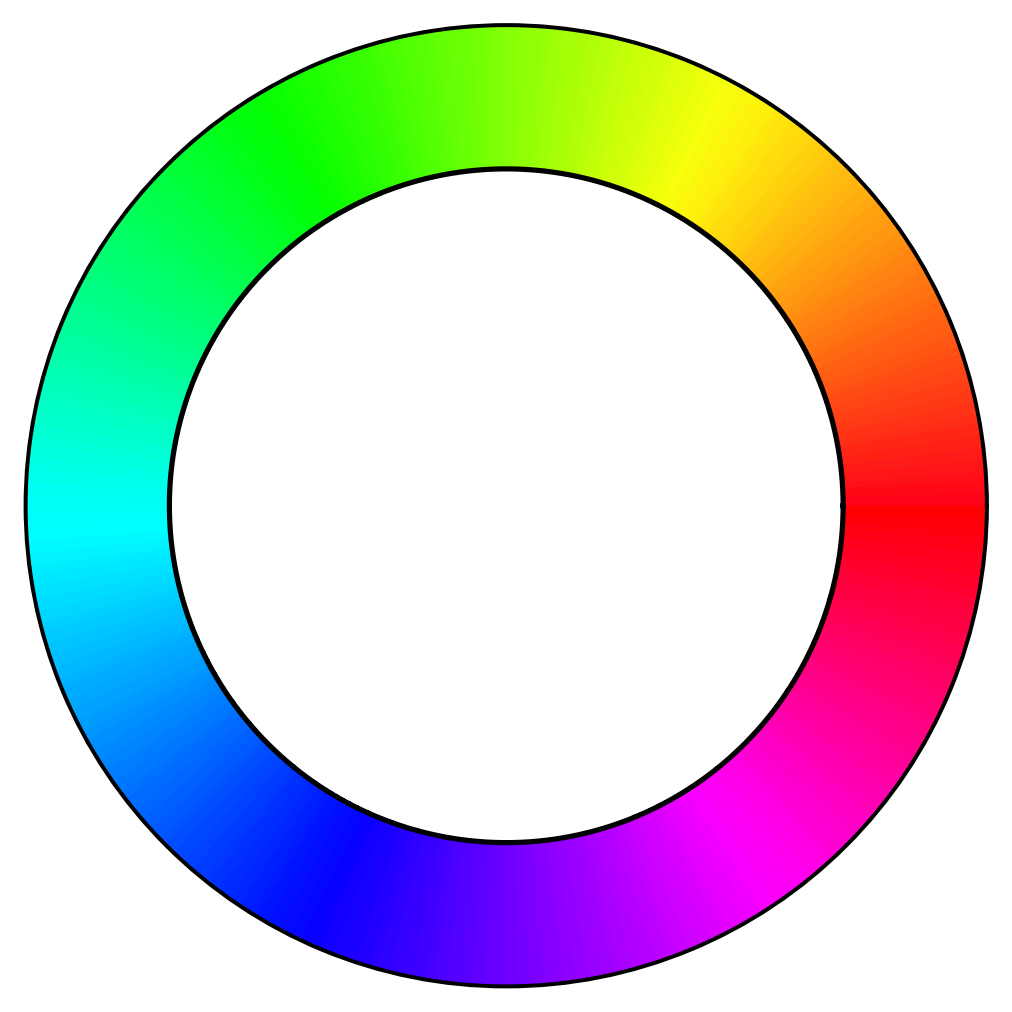}%
        };
    \end{tikzpicture}

    \caption{Simulation snapshots, centered on the obstacle, with membranes colored according to their velocity orientation: angle between the velocity and the $x$ axis. The mapping of angles to colors is illustrated in the color wheel located between the figures.}
    \label{fig:snapshots_color_vel}
\end{figure*}

\section{Exploration method}
\label{sec:exploration}
Following the standard procedures used in the work of Beatrici \textit{et al.} \cite{beatrici_comparing_2023}, we selected three physical quantities of interest and associated each one with a model parameter expected to control it:
\begin{itemize}
    \item Velocity alignment (A): associated with the relaxation time $\tau$ in \cref{eq:eq_motion}.  
    \item Adhesion between membranes (T): associated with $k_\mathrm{adh}$ in \cref{eq:energy_c_adh};
    \item Density (D): associated with the force $F_i$ pushing the membranes at the creation region.  
    
\end{itemize}
We ran simulations for all combinations of the selected parameter extremes. The parameter ranges are constrained by numerical limitations and physical considerations. For instance, if $F_i$ is too large, numerical errors will dominate the simulations. Therefore, there are $2^3=8$ possible scenarios to simulate. Each simulation is defined by the selected parameter values, with each parameter taking one of two possible values: high or low. To organize the simulations, we assign each parameter a letter reflecting the physical property it influences (see \cref{tab:parameter_summary} for details). We then use the numbers $0$ or $1$ to indicate whether the parameter is set to the value that tends to minimize ($0$) or maximize ($1$) the associated physical quantity, namely, A, for velocity alignment, T, for adhesion between membranes and D, for cell density.  Thus, each simulation is labeled as follows:
\begin{center}
  A[0 or 1]-T[0 or 1]-D[0 or 1]
\end{center}
For example, the \case{1}{0}{1} corresponds to a simulation where $\tau$ has a low value ($A1$), since lower $\tau$ promotes more collective motion, $k_\mathrm{adh}$ has a high value ($T1$), and $F_i$ has a low value ($D0$).

The extreme values chosen for $\tau$ are $0.5\tau_0$ and $25\tau_0$. The adhesion elastic constant $k_\mathrm{adh}$ was varied between $0.556 \epsilon/\sigma^2$ and $33.333 \epsilon/\sigma^2$, ensuring finite adhesion while avoiding excessively large values that would require smaller integration time steps. Finally, the force controlling density was varied between $v_0/\mu$ and $10v_0/\mu$, where $v_0/\mu$ sets the characteristic active force scale. The chosen parameter values are summarized in \cref{tab:parameter_summary}.

\begin{table}[h!]
    \centering
    \small
    \caption{Mapping between the explored parameters, simulation labels,
    parameter ranges, and associated physical quantities. The label letter is chosen to reflect the physical quantity to which the parameter is related}
    \label{tab:parameter_summary}

    \begin{tabular*}{\linewidth}{
        @{\extracolsep{\fill}}
        lllll
    }
        \hline
        Parameter
        & Label
        & Min. value
        & Max. value
        & Associated quantity \\
        \hline

        $\tau$
        & A
        & $0.5\,\tau_0$
        & $25\,\tau_0$
        & Alignment [\cref{eq:input_alignment}] \\

        $k_\mathrm{adh}$
        & T
        & $0.556\,\epsilon/\sigma^2$
        & $33.333\,\epsilon/\sigma^2$
        & Adhesion / Tension \\

        $F_i$
        & D
        & $1\,v_0/\mu$
        & $10\,v_0/\mu$
        & Density [\cref{eq:rel_den}] \\

        \hline
    \end{tabular*}
\end{table}

\section{Measurements}
\label{sec:measuments}
During the simulations, two types of measurements are performed: input and output measurements. These measurements are taken after the number of membranes in the channel stabilizes, which occurs at approximately $10^4$ membranes. We also perform temporal averaging to improve the statistics.

\begin{figure*}[h!]
    \centering
    \begin{subfigure}{0.49\textwidth}
        \centering
        \includegraphics[width=\textwidth]{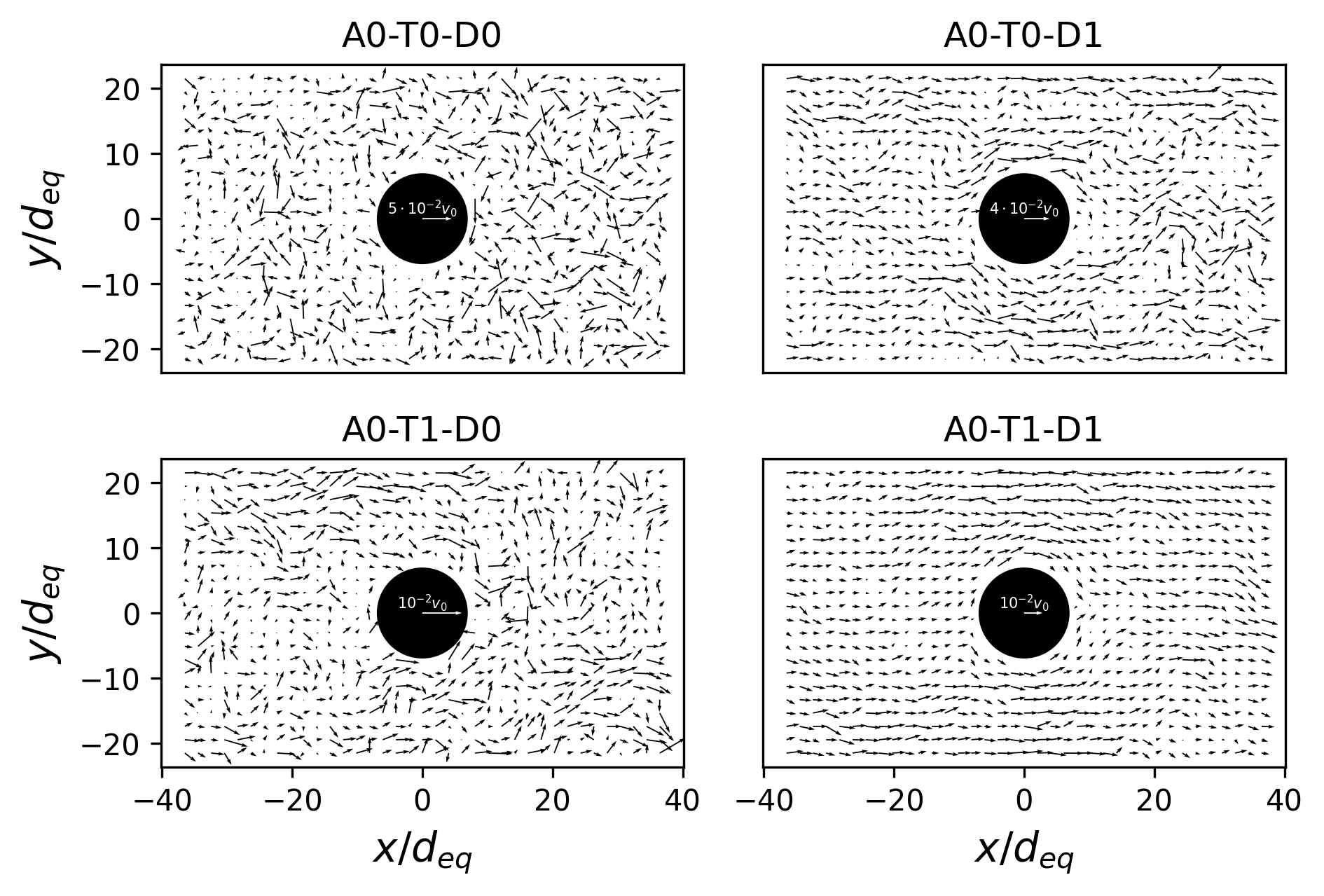}
        \caption{High $\tau$}
        \label{fig:low_align_output_vel}
    \end{subfigure}
    \hfill
    \begin{subfigure}{0.49\textwidth}
        \centering
        \includegraphics[width=\textwidth]{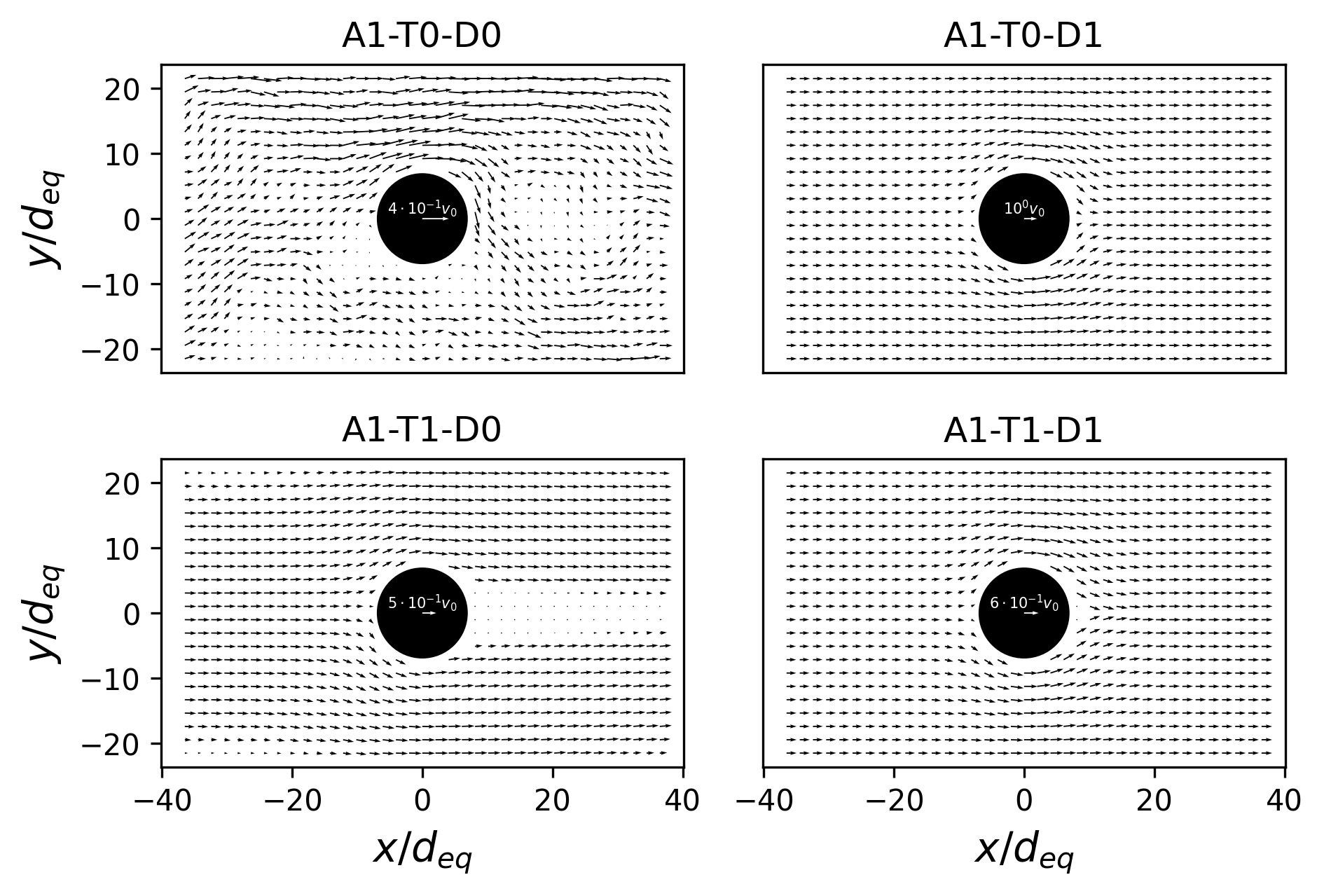}
        \caption{Low $\tau$}
        \label{fig:high_align_output_vel}
    \end{subfigure}
    
    \caption{Velocity fields for all simulated cases. The black circle represents the channel obstacle. The white arrow inside it is a reference velocity vector, with its magnitude indicated above; the arrow scale differs between graphs. Panel \subref{fig:low_align_output_vel} shows the four cases with low alignment and panel \subref{fig:high_align_output_vel} the four cases with high alignment. It is clear that the velocity field is significantly more organized in cases with high alignment compared to those with low alignment and the field is usually symmetric with respect to the $x$-axis, with the only exception being the \case{1}{0}{0}.}
    \label{fig:vel_output_measurements}
\end{figure*}

\begin{figure*}[h!]
    \centering
    \begin{subfigure}{0.49\textwidth}
        \centering
        \includegraphics[width=\textwidth]{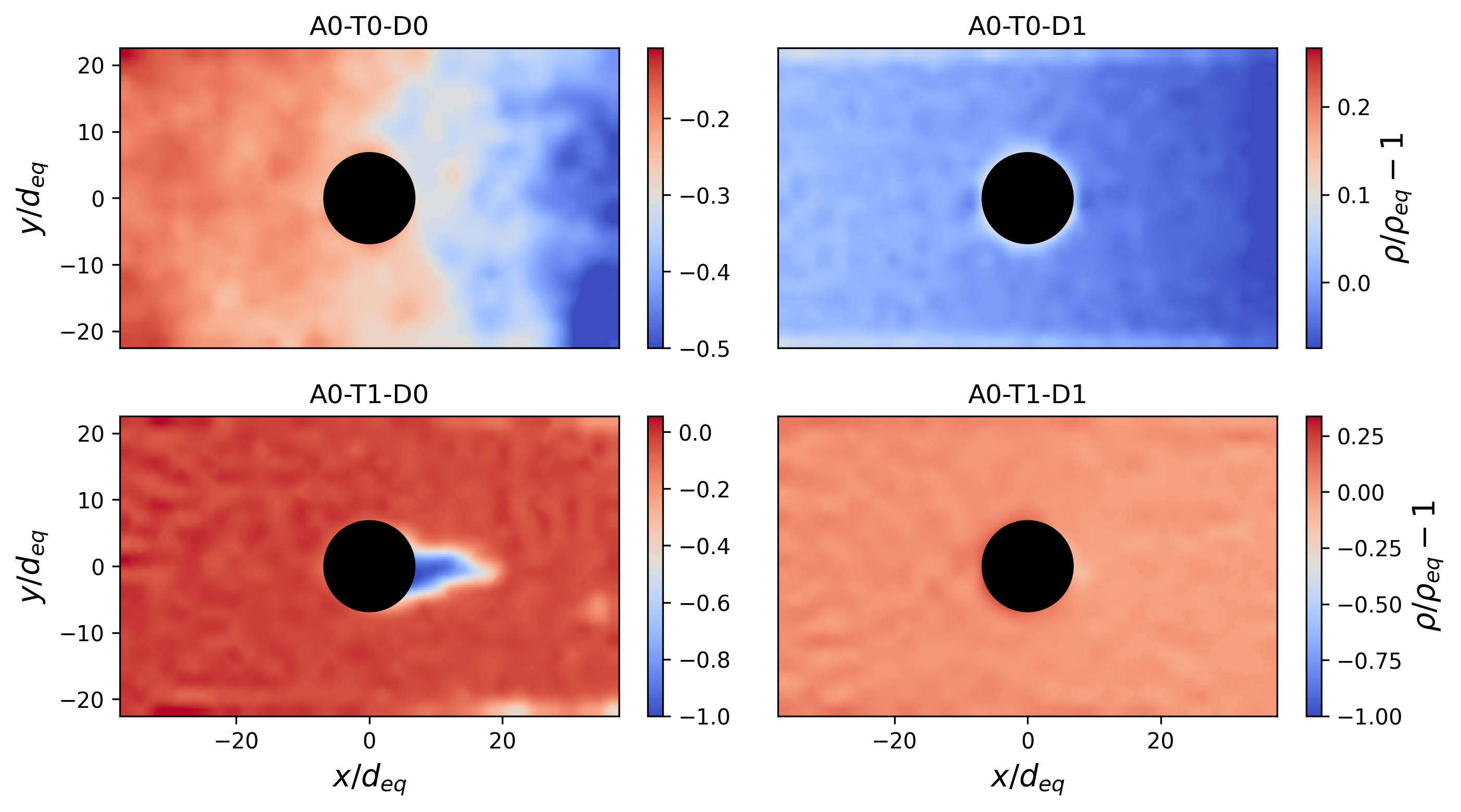}
        \caption{High $\tau$}
        \label{fig:low_align_output_den}
    \end{subfigure}
    \hfill
    \begin{subfigure}{0.49\textwidth}
        \centering
        \includegraphics[width=\textwidth]{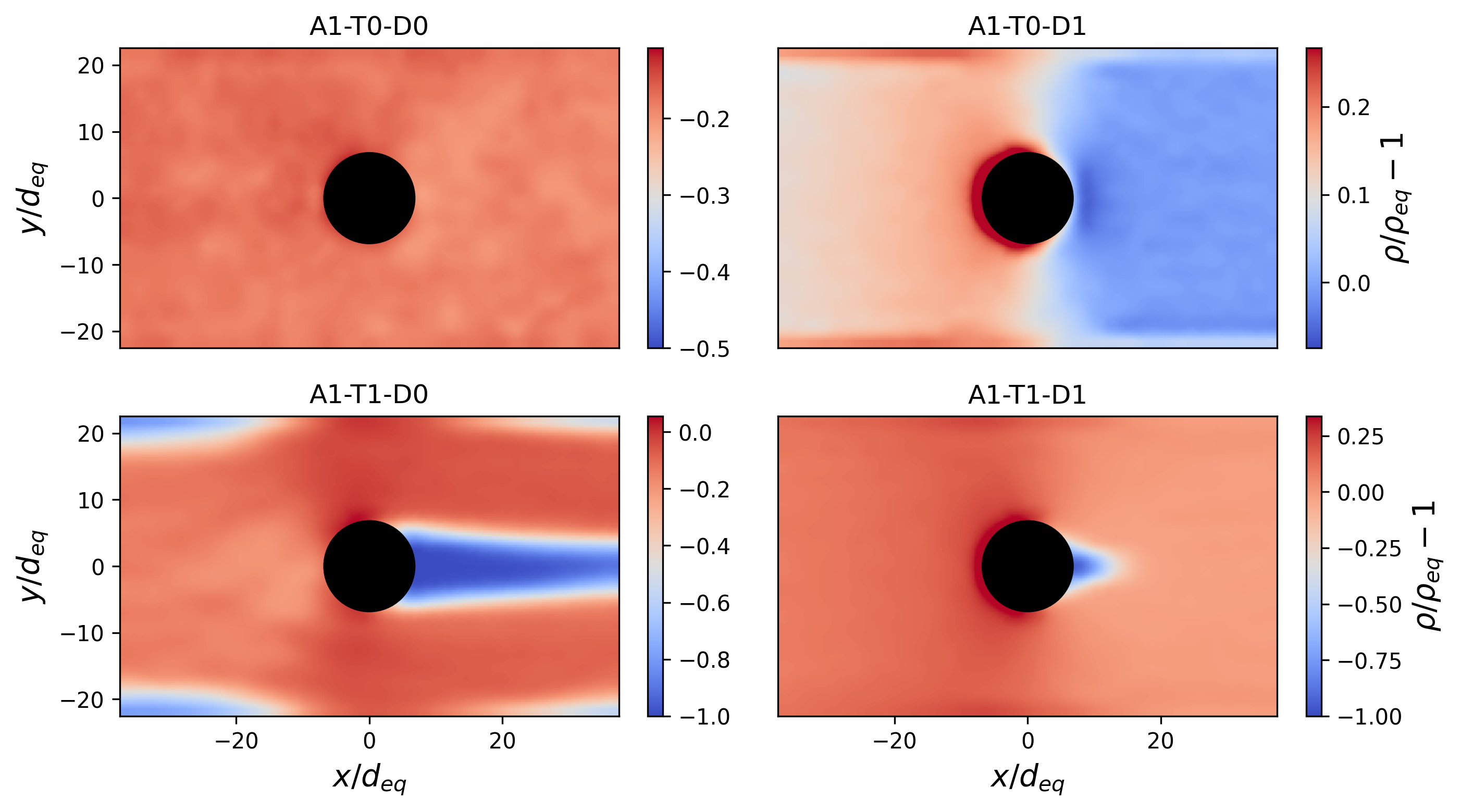}
        \caption{Low $\tau$}
        \label{fig:high_align_output_den}
    \end{subfigure}
    
    \caption{Relative density fields for all simulated cases shown as color maps with values increasing from blue to red. Panel \subref{fig:low_align_output_den} shows the four cases with low alignment and panel \subref{fig:high_align_output_den} the four cases with high alignment. For each panel, the color scale on the right side of each graph indicates the specific scale used, which may differ between graphs. However, graphs in the same position across different panels share the same scale, allowing for easy comparison in cases where only the alignment changes. Note that the lowest value the relative density can have is $-1$, which corresponds to a complete vacuum. The black circle is the channel obstacle. The overall effect of the obstacle is a reduction in density downstream of its location, as indicated by the predominance of blue tones on the right side of the graphs compared to the left. In cases with high adhesion (cases with $T1$ in their labels, i.e., the bottom row in each panel), a vacuum region appears behind the obstacle. The size of this region is influenced by the density parameter, becoming smaller for higher density values (when the case has $D1$ in its label).}
    \label{fig:den_output_measurements}
\end{figure*}

\subsection{Input measurements}
Input measurements are dimensionless measurements that monitor the entrance of the simulation region 
and serve as a set of numbers that can be used to compare different models in the Stokes geometry. 
These measurements are taken within the input region (yellow region in \cref{fig:channel_view}), a place
that is sufficiently far away from the beginning of the channel and the obstacle, because, ideally, this set of
measurements should not be significantly affected by the obstacle or the membrane-creation region. 
The following quantities compose the input measurements:

\begin{itemize}
    \item Velocity Alignment ($\Psi$) \cite{vicsek_novel_1995}: Measures how well the velocities of the membranes are aligned
    \begin{equation}
    \Psi = \frac{1}{N_{\mathrm{input}}} \norm{\sum_{j=1}^{N_{\mathrm{input}}} \frac{\mathbf{v}_j}{|\mathbf{v}_j|}}
    \label{eq:input_alignment}
    \end{equation}
    where $\mathbf{v}_j$ denotes the center-of-mass velocity of membrane $j$, and $N_{\mathrm{input}}$ is the number of membranes whose geometric centers lie within the input region. Values of $\Psi \approx 0$ indicate disordered flow, whereas $\Psi \approx 1$ corresponds to strongly aligned, highly ordered flow.
    \item Relative Density ($\delta \rho$): Density relative to an equilibrium value $\rho_{eq}  = 1 / A_{0}$
    \begin{equation}
    \delta \rho = \frac{\rho}{\rho_{eq}} - 1
    \label{eq:rel_den}
    \end{equation}
    where $\rho$ is the local density, defined as the number of membranes per unit area.
    \item Solid--Liquid Behavior \cite{gregoire_moving_2003} ($\Delta$): A measurement indicating whether the ensemble of membranes has liquid or solid behavior. Suppose that there are $N_{\mathrm{input}}$ membranes inside the input region
    at time $t$, and that the $i$-th membrane has $n_i$ neighbors. Membranes are neighbors if they are
    neighbors in the Voronoi diagram created by their geometric centers, and their distance is smaller than $1.4d_{eq}$. Let $B_i$ be the set 
    of neighbors of membrane $i$ at time $t$ and $r_{ij}(t)$ the distance between membranes $i$ and $j$
    at time $t$ (measured using their geometric centers), then the contribution to $\Delta$ from membrane $i$ 
    is given by \cref{eq:delta_i}
    \begin{equation}
    \Delta_i = 1 - \frac{1}{n_i}\sum_{j \in B_i}\bigg(1 - \frac{r_{ij}^2(t)}{r_{ij}^2(t + T)} \bigg)
    \label{eq:delta_i}
    \end{equation}
    where $T$ is the time required for the geometric center of the $N_{\mathrm{input}}$ membranes to have moved by 
    one obstacle radius. Finally, $\Delta$ is defined as
    \begin{equation}
        \Delta = \frac{1}{N_{\mathrm{input}}}\sum_{i=1}^{N_{\mathrm{input}}}\Delta_i
        \label{eq:delta}
    \end{equation}
Intuitively, this quantity indicates whether neighbors are exchanged, as in liquid-like behavior ($\Delta \approx 0$), or remain unchanged, as in solid-like behavior ($\Delta \approx 1$).
\end{itemize}

\subsection{Output measurements}
Output measurements consist of spatial fields evaluated in the region surrounding the obstacle (purple region in \cref{fig:channel_view}). The following quantities are measured:
\begin{itemize}
    \item Relative Density Field ($\delta \rho$): The same quantity as defined in the input measurements, but in this case the result is a scalar spatial field (over the output region), not a single number.

    \item Velocity Field ($\mathbf v$): Coarse-grained velocity field obtained from the center-of-mass velocities of the membranes.
\end{itemize}

The spatial fields are computed by coarse-graining the data over rectangular windows of side length $2d_{eq}$. For more details, see \Cref{app:coarse_grainig} of the SI.

\section{Simulation results}
We now present the simulation results, beginning with the input measurements and then considering the output measurements.
In what follows, results from the eight simulated cases are divided into two sets of four: one set contains all four cases with high $\tau$ (cases with $A0$ in their labels, usually related to low collective motion) and the other contains all four cases with low $\tau$ (cases with $A1$ in their label, usually related to more collective motion). \Cref{fig:snapshots_color_vel} shows steady-state snapshots for all simulated cases, while Supplementary Movies \ref{mov:A0T0D0}A--\ref{mov:A1T1D1}A show the corresponding full-channel simulations and Supplementary Movies \ref{mov:A0T0D0}B--\ref{mov:A1T1D1}B provide enlarged views of the region around the obstacle.

\subsection{Input measurements}
The input measurements, compared with those of the Multiparticle Model (MPM) explored in our reference work~\cite{beatrici_comparing_2023}, are shown in \cref{tab:all_input_measurements}. The $\Psi$ column demonstrates that, except for low densities, the DAMM achieves higher collective motion than the MPM. Averaging the $\Psi$ values for cases \casenl{1}{1}{0}, \casenl{1}{0}{1}, and \casenl{1}{1}{1} (the last three rows in \cref{tab:all_input_measurements}), the DAMM yields $\Psi=0.96$, which is significantly greater than the MPM value of $\Psi=0.69$.

The relative density, $\delta\rho$, under high and low $F_i$ conditions reveals physically consistent behavior in the DAMM. The $D1$ cases exhibit systematically higher relative densities than their $D0$ counterparts. Because $\delta\rho=0$ corresponds to the equilibrium density, the $D1$ cases yield positive values, whereas the $D0$ cases yield negative ones. This consistent behavior is not observed in the MPM. For example, \case{0}{1}{0} has a lower $\delta \rho$ than \case{0}{0}{0}, and almost all cases with $D1$ have negative $\delta \rho$ values.

As a final point, the DAMM exhibits both liquid-like ($\Delta \approx 0$) and solid-like ($\Delta \approx 1$) behavior, with minimum and maximum $\Delta$ values of $0.02$ and $1.02$ observed in cases \casenl{0}{0}{0} and \casenl{1}{1}{1}, respectively. In contrast, the MPM showed minimum and maximum values of $0.09$ and $0.72$ in cases \casenl{0}{0}{0} and \casenl{1}{0}{1}, respectively. This indicates that the DAMM was able to reach values closer to the extremes of the solid--liquid spectrum.

    

\begin{table}[h!]
\centering
\caption{Input measurements for the eight simulated cases, comparing the MPM and the DAMM. Cases are grouped by the alignment parameter $\tau$: the first four rows correspond to low alignment, and the last four to high alignment. Each input measurement—alignment ($\Psi$), relative density ($\delta \rho$), and solid--liquid behavior ($\Delta$)—is presented in a column with two subcolumns, one for each model. Compared to the MPM, the DAMM achieves higher alignment, displays more consistent relative density values, and reaches values closer to the extremes of the solid--liquid spectrum, as discussed in the text}

\begin{tabular*}{0.48\textwidth}{@{\extracolsep{\fill}}c|cc|cc|cc}
\hline
Case & \multicolumn{2}{c|}{$\Psi$} & \multicolumn{2}{c|}{$\delta \rho$} & \multicolumn{2}{c}{$\Delta$} \\
A-T-D & MPM & DAMM & MPM & DAMM & MPM & DAMM \\
\hline
000 & 0.182 & 0.162 & 0.098 & -0.156 & 0.086 & 0.016 \\
001 & 0.161 & 0.190 & -0.013 & 0.031 & 0.145 & 0.191 \\
010 & 0.285 & 0.234 & -0.050 & -0.021 & 0.337 & 0.955 \\
011 & 0.290 & 0.218 & -0.010 & 0.045 & 0.353 & 1.006 \\
\hline
100 & 0.643 & 0.300 & -0.028 & -0.170 & 0.439 & 0.136 \\
101 & 0.599 & 0.960 & 0.030 & 0.112 & 0.654 & 1.013 \\
110 & 0.671 & 0.948 & -0.240 & -0.261 & 0.720 & 1.002 \\
111 & 0.794 & 0.968 & -0.079 & 0.092 & 0.708 & 1.015 \\
\hline
\end{tabular*}
\label{tab:all_input_measurements}
\end{table}


\subsection{Output measurements}
In this section we analyze the output measurements, focusing on the relative density and velocity fields to characterize the behavior of the simulated tissue around the obstacle.

Velocity fields with low $\tau$ (\cref{fig:high_align_output_vel}) exhibit a well-ordered field, as anticipated, and display strong symmetry with respect to a horizontal axis passing through the center of the obstacle, with the only exception being the \case{1}{0}{0}, where cells bypass the obstacle faster at the top than at the bottom. In these cases, the typical membrane speed is approximately $v_0/2$, indicating that the presence of the obstacle tends to reduce the collective motion, with the exception being the \case{1}{1}{0}, where cells flow around the obstacle at a speed of approximately $v_0$. On the other hand, cases with high $\tau$ ($A0$) exhibit fields with less cohesive structure, although \cref{fig:low_align_output_vel} shows that high $F_i$ ($D1$) can still induce some degree of collective motion. The typical membrane speed is approximately $10^{-2}v_0$, nearly two orders of magnitude lower than that observed at smaller $\tau$ ($A1$), consistent with the role of $\tau$ in controlling collective motion.

\begin{figure*}[ht!]
    \centering

    \begin{tikzpicture}[
        image/.style={
            inner sep=0,
            outer sep=0
        },
        connection/.style={
            -{Latex[length=3mm,width=2mm]},
            thick
        },
        panel label/.style={
            anchor=north west,
            font=\bfseries,
            fill=white,
            fill opacity=0.7,
            text opacity=1,
            inner sep=1.5pt,
            rounded corners=1pt
        }
    ]


        \node[image] (figA)
        {
            \includegraphics[
                width=0.42\textwidth
            ]{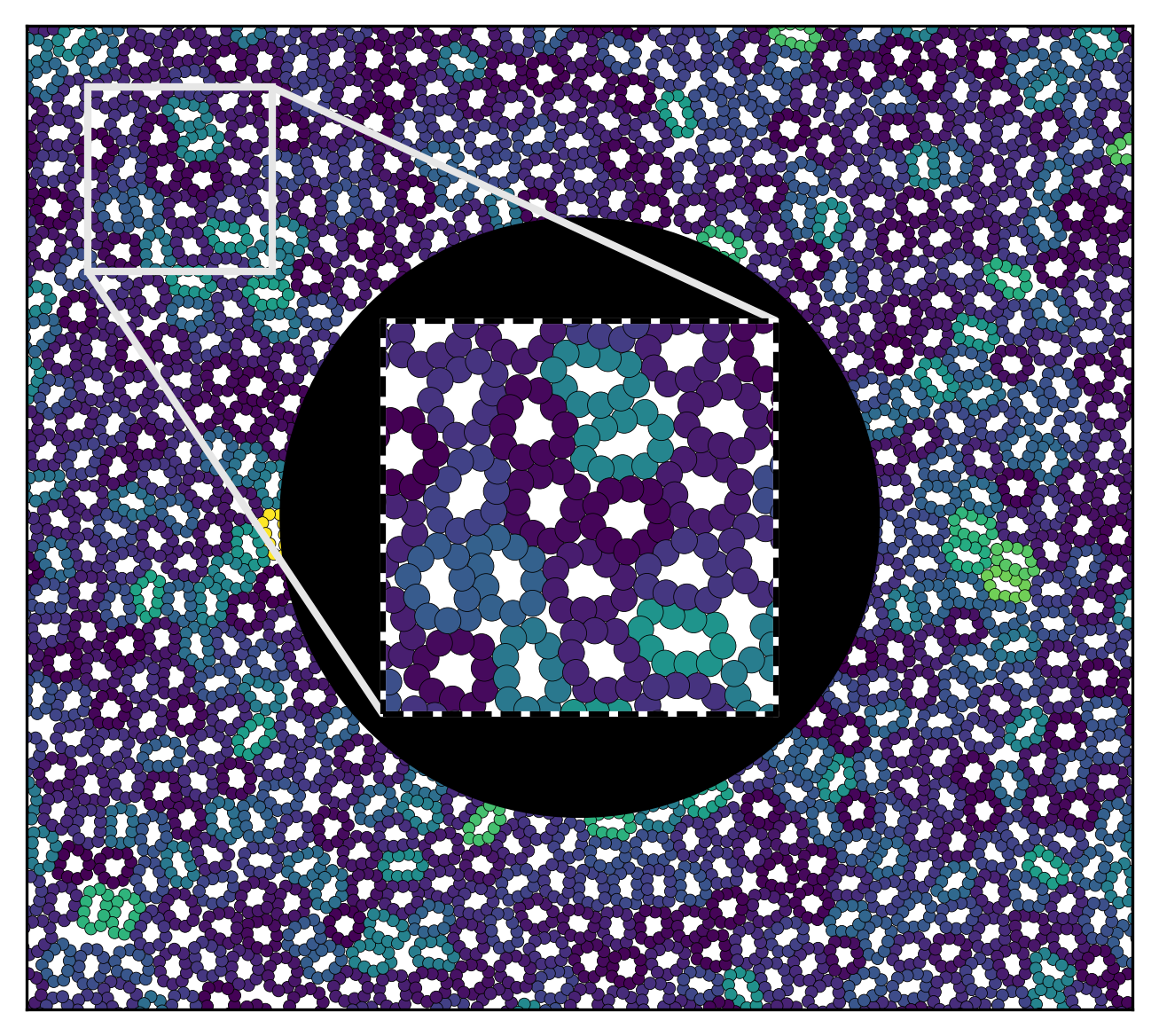}
        };

        \node[
            image,
            right=4mm of figA
        ] (figB)
        {
            \includegraphics[
                width=0.42\textwidth
            ]{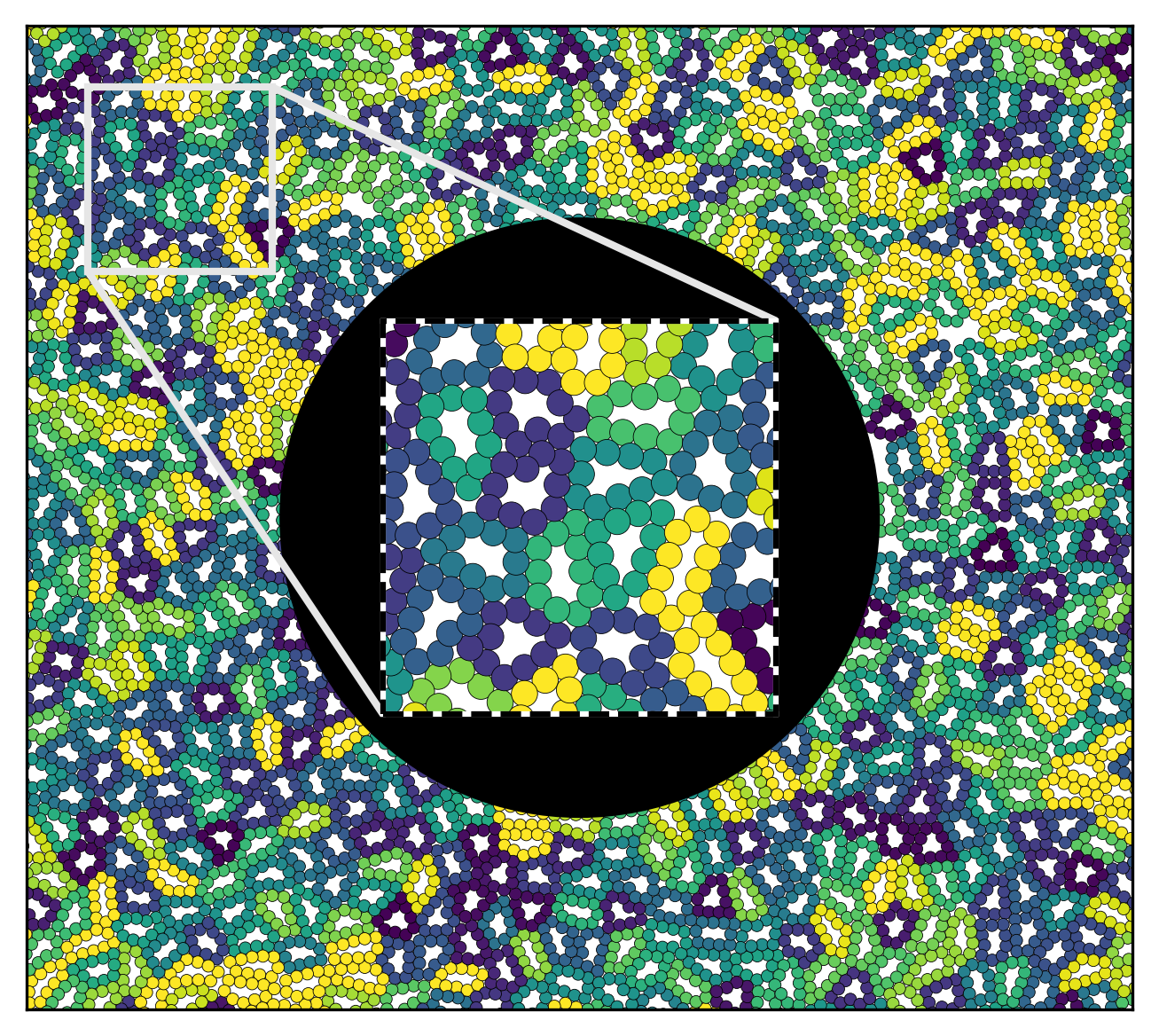}
        };

        \node[
            image,
            right=2mm of figB,
            anchor=west
        ] (colorbar)
        {
            \includegraphics[
                width=0.06\textwidth
            ]{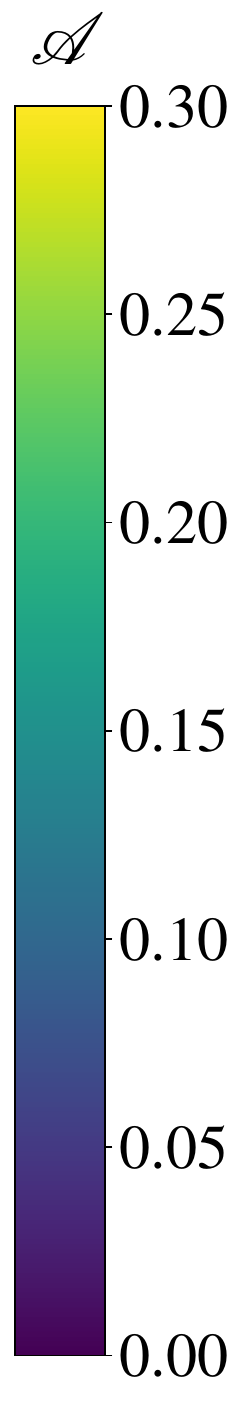}
        };


        \node[
            image,
            below=5mm of figA
        ] (figC)
        {
            \includegraphics[
                width=0.42\textwidth
            ]{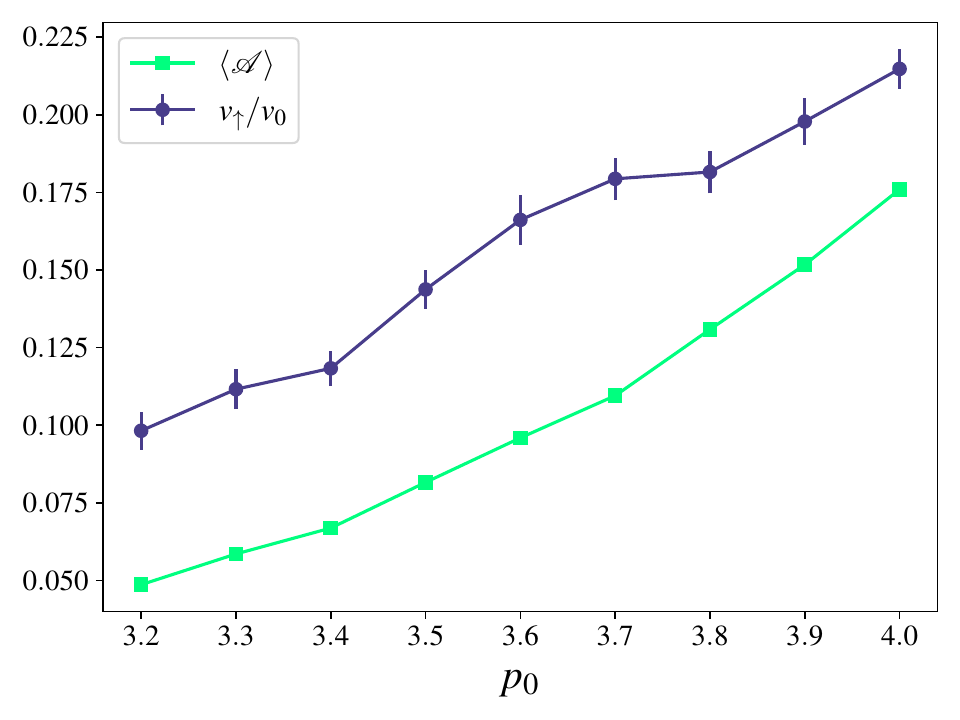}
        };

        \node[
            image,
            right=4mm of figC
        ] (figD)
        {
            \includegraphics[
                width=0.42\textwidth
            ]{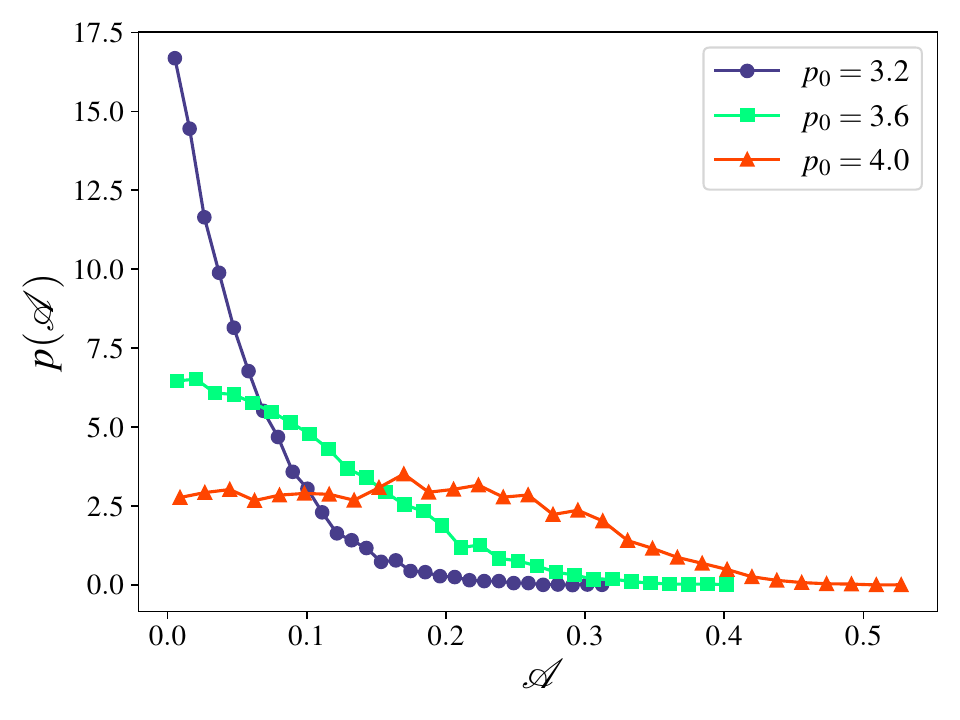}
        };


        \node[panel label]
            at ($(figA.north west)+(1mm,-1mm)$)
            {(a)};

        \node[panel label]
            at ($(figB.north west)+(1mm,-1mm)$)
            {(b)};

        \node[panel label]
            at ($(figC.north west)+(1mm,3mm)$)
            {(c)};

        \node[panel label]
            at ($(figD.north west)+(1mm,3mm)$)
            {(d)};


        \begin{scope}[
            shift={(figC.south west)},
            x={($(figC.south east)-(figC.south west)$)},
            y={($(figC.north west)-(figC.south west)$)}
        ]

            \coordinate (firstPoint) at (0.18,0.75);
            \coordinate (lastPoint)  at (0.96,0.95);


        \end{scope}


        \node[
            anchor=south west,
            font=\bfseries
        ]
        at ($(firstPoint)+(-2mm,-14mm)$)
        {Stiff};

        \node[
            anchor=north east,
            font=\bfseries
        ]
        at ($(lastPoint)+(1mm,-24mm)$)
        {Flexible};


        \node[
            anchor=south west,
            font=\bfseries
        ]
        at ($(figD.north west)+(18mm,-10mm)$)
        {Circle-Like};

        \node[
            anchor=south east,
            font=\bfseries
        ]
        at ($(figD.south east)+(-2mm,13mm)$)
        {Elongated};




    \end{tikzpicture}

    \caption{Exploration of the shape index $p_0$. Panels (a, b): steady-state simulation snapshots centered on the channel obstacle for $p_0=3.2$ and $p_0=4.0$, respectively. Cells are rounder and more rigid at $p_0=3.2$, whereas they are considerably more deformable at $p_0=4.0$. Cell colors indicate asphericity ($\mathcal{A}$). The inset within the obstacle shows a magnified view of a selected square region of the tissue. Panel (c): effect of cell deformability parameter $p_0$ on tissue velocity ($v_{\uparrow}$) and mean asphericity over all membranes in the channel ($\langle \mathcal{A} \rangle$). Both $v_{\uparrow}$ and $\langle \mathcal{A} \rangle$ increase monotonically with increasing $p_0$, demonstrating that more flexible cells enable faster collective migration through the channel. Error bars represent the standard error of the mean. Panel (d): asphericity distribution ($p(\mathcal{A})$) for all membranes in the channel at the final state, shown for three values of $p_0$. At lower $p_0$, the membranes are predominantly circular, resulting in a pronounced peak at $\mathcal{A}=0$. As $p_0$ increases, the distribution broadens toward larger $\mathcal{A}$, indicating increasingly elongated membrane shapes.}
    \label{fig:snapshots_simulacao}

    \refstepcounter{subfigure}
    \label{fig:small_po}
    \edef\labelA{\thesubfigure}
    
    \refstepcounter{subfigure}
    \label{fig:big_po}
    \edef\labelB{\thesubfigure}
    
    \refstepcounter{subfigure}
    \label{fig:po_exploration}
    \edef\labelC{\thesubfigure}
    
    \refstepcounter{subfigure}
    \label{fig:pa_exploration}
    \edef\labelD{\thesubfigure}
\end{figure*}
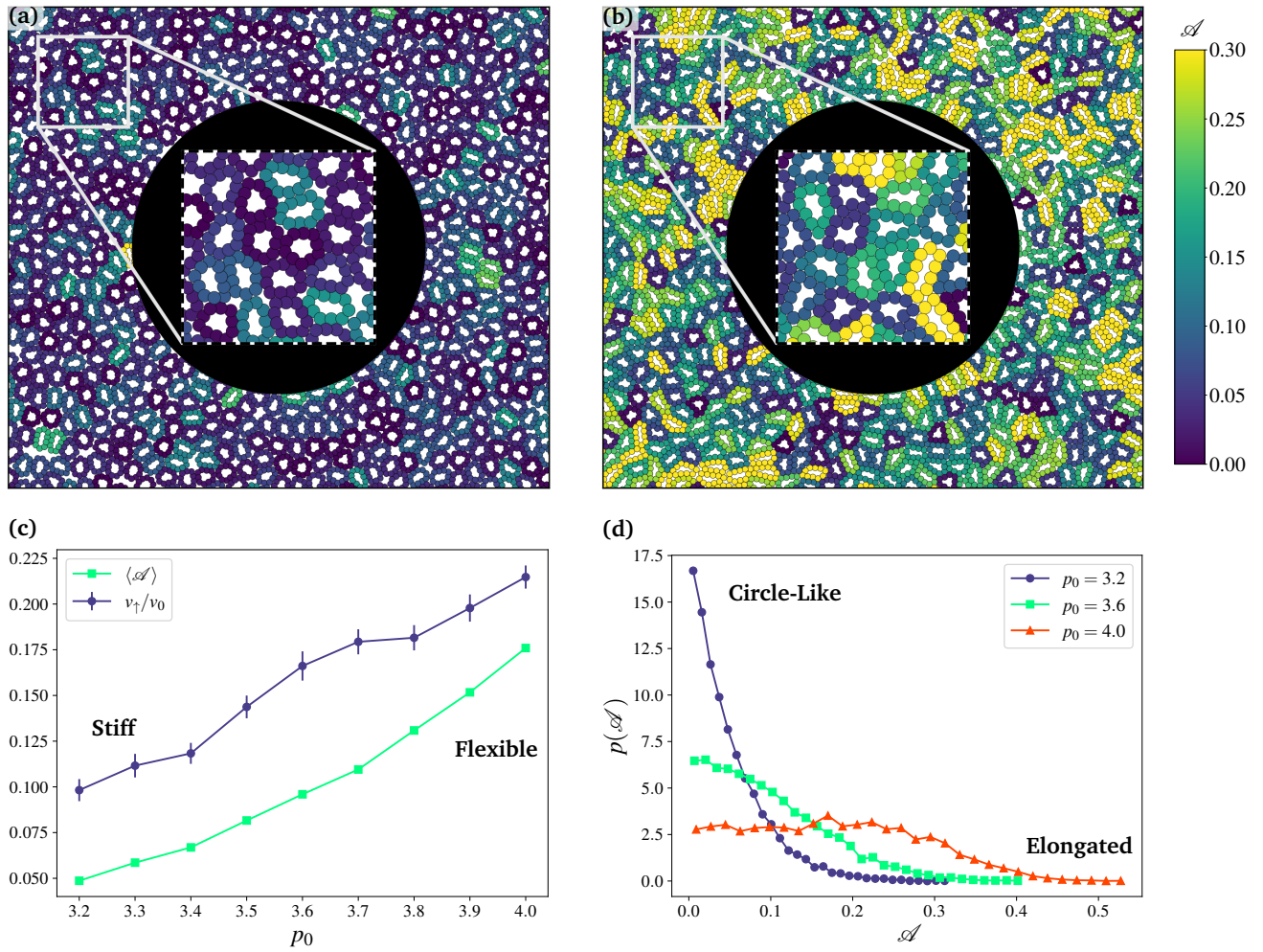

Relative density fields are represented by color maps with values 
increasing from blue to red. The overall effect of the 
obstacle is to decrease the density downstream; this can be seen in
\cref{fig:den_output_measurements}, noting that the right half of the graphs
is bluer than the left half. The effects of adhesion are particularly 
evident: in cases with high adhesion ($T1$), voids appear just after the obstacle.
This void region becomes smaller when the density of the membranes is also 
high ($D1$), which indicates a partial compensation of the obstacle 
effect by the greater amount of material. A notable example occurs in \case{1}{0}{1}, in which the void region extends to the end of the channel.
Finally, it is worth mentioning that density field fluctuations in the cases with higher $\tau$ (\cref{fig:low_align_output_den})
are greater than those in the cases with lower $\tau$ (\cref{fig:high_align_output_den}), consistent with 
the results obtained from velocity fields.

Although a direct comparison between the DAMM and the full set of models examined in Ref.~\cite{beatrici_comparing_2023} lies beyond the scope of this study, the MPM provides a particularly suitable benchmark because both belong to the same class of deformable-particle models. Their comparison is also informative because they preserve cell area through distinct mechanisms: the DAMM employs a simple area-elasticity term, whereas the MPM represents the cell interior through a nucleus connected by springs to the boundary particles. The two models further differ in their treatment of cell motility. In the DAMM, the polarity responds to the cell velocity through a self-alignment feedback mechanism, providing a more biologically motivated description than the Vicsek-like alignment rule adopted in the MPM. Consequently, the performance and behavior of the DAMM can be transitively compared to the remaining four models (Vicsek model~\cite{vicsek_novel_1995,gregoire_moving_2003}, Self-aligning elastic-disk model~\cite{szabo_phase_2006}, Voronoi model~\cite{barton_active_2017} and Potts model~\cite{kafer_moving_2006,kabla_collective_2012}) from the reference work, establishing a clear context for our findings within the broader literature. 

The comparison also highlights important similarities and differences between the DAMM and the MPM. In terms of collective alignment, the DAMM reaches higher maximum alignment values than the MPM, approaching those observed in particle-based models such as the self-aligning elastic-disk model~\cite{beatrici_comparing_2023}. In this respect, the DAMM reproduces both the individual and collective behavior of self-aligning elastic disks while retaining the crucial ability to deform and change their shape in response to mechanical interactions, allowing the model to capture cell-shape adaptations that are essential in biological tissues.

In contrast, the range of relative densities observed in the DAMM is similar to that reported for the MPM and remains narrower than that obtained with particle-based models. This limitation is associated with the finite compressibility of the membranes, which constrains the maximum density that can be achieved in highly crowded regions. In principle, larger density variations could be obtained by increasing $k_P$ and $k_c$, thereby allowing cells to sustain higher internal pressures. However, stiffer membranes introduce faster dynamical timescales, requiring smaller integration time steps for numerical stability and consequently leading to substantially higher computational costs.

\section{Membrane deformability promotes shape elongation and faster tissue flow}
\label{sec:po_exploration}

To investigate how membrane deformability affects tissue migration, we varied the target shape index from $p_0=3.2$ to $p_0=4.0$, while keeping the target area $A_0$ fixed at the value reported in \cref{tab:fix_pars}. The shape index was varied by changing the target bond length $l_0$. Representative configurations of tissues composed of less and more deformable membranes are shown in \cref{fig:small_po,fig:big_po}, respectively.

The investigated range spans the characteristic value $p_0^\ast \simeq 3.81$. In confluent vertex and Voronoi models, this value marks a rigidity transition from a solid-like state for $p_0<p_0^\ast$ to a liquid-like state for $p_0>p_0^\ast$~\cite{bi_density-independent_2015,bi_motility-driven_2016,barton_active_2017}. A similar characteristic value is associated with the jamming transition in the deformable-particle model of Boromand \textit{et al.}~\cite{PhysRevLett.121.248003}. We therefore consider values on both sides of $p_0^\ast$ to examine how increasing membrane deformability influences tissue migration through the channel.

Tissue mobility was quantified by the mean velocity along the channel direction in the input region, denoted by $v_{\uparrow}$. This quantity was measured after the number of membranes inside the channel had reached a statistically stationary value. Specifically, $v_{\uparrow}$ was calculated as the time-averaged $x$-component of the mean center-of-mass velocity of all membranes whose geometric centers were located within the input region.

To quantify the accompanying changes in membrane morphology, we used the gyration tensor $\Re_{j}(t)$~\cite{paoluzzi2016shape,tian2017anomalous,wang2019shape}, defined for each membrane as
\begin{equation}
\label{tensor_gir}
\boldsymbol{\Re}_j(t)
=
\frac{1}{n}
\sum_{i=1}^{n}
\mathbf{r}_{i,\mathrm{CM}}^{\,j}(t)
\otimes
\mathbf{r}_{i,\mathrm{CM}}^{\,j}(t),
\end{equation}
where $n$ is the number of particles composing membrane $j$,
$\mathbf{r}_{i,\mathrm{CM}}^{\,j}(t)
=
\mathbf{r}_{i,j}(t)-\mathbf{r}_{\mathrm{CM},j}(t)$
is the position of particle $i$ relative to the center of mass of membrane $j$, and $\otimes$ denotes the tensor product. In two dimensions, the gyration tensor can be written as
\begin{equation}
\label{comp_gir}
\boldsymbol{\Re}_j(t)
=
\begin{bmatrix}
R_{xx}^{\,j}(t) & R_{xy}^{\,j}(t) \\
R_{yx}^{\,j}(t) & R_{yy}^{\,j}(t)
\end{bmatrix}.
\end{equation}

The elongation of membrane $j$ was quantified by its asphericity~\cite{paoluzzi2016shape,tian2017anomalous},
\begin{equation}
\label{asphericity}
\mathcal{A}_j(t)
=
\frac{
\left[\lambda_1^{\,j}(t)-\lambda_2^{\,j}(t)\right]^2
}{
\left[\lambda_1^{\,j}(t)+\lambda_2^{\,j}(t)\right]^2
},
\end{equation}
where $\lambda_1^{\,j}(t)$ and $\lambda_2^{\,j}(t)$ are the eigenvalues of the gyration tensor $\boldsymbol{\Re}_j(t)$. The asphericity ranges from $\mathcal{A}_j=0$ for an isotropic shape, such as a circle, to $\mathcal{A}_j\rightarrow1$ for a highly elongated, rod-like configuration.

The simulations were performed with $\tau=1.5\tau_0$, $k_{\mathrm{adh}}=2.778\,\epsilon/\sigma^2$, and $F_i=4v_0/\mu$, with all remaining parameters listed in \cref{tab:fix_pars}. Based on the preceding parameter exploration, these values were selected to produce a nearly confluent tissue with relatively weak global alignment, allowing the influence of membrane deformability to be examined.

As shown in \cref{fig:po_exploration}, both the mean flow speed $v_{\uparrow}$ and the mean membrane asphericity $\langle\mathcal{A}\rangle_j$ increase with $p_0$. Thus, the systems with more deformable membranes flow faster through the channel and exhibit more elongated membrane shapes. \Cref{fig:pa_exploration} shows the asphericity distribution over all membranes in the final configuration of each simulation. For $p_0=3.2$, the distribution exhibits a pronounced peak near $\mathcal{A}=0$, indicating that most membranes remain approximately circular. As $p_0$ increases, the distribution broadens and shifts toward larger values of $\mathcal{A}$, reflecting increasingly elongated and heterogeneous membrane shapes. These results indicate that greater membrane deformability enables larger shape changes and facilitates local rearrangements, thereby promoting faster tissue flow through the channel.
\begin{figure*}[h!]
    \centering
      \begin{subfigure}[t]{0.49\linewidth}
        \includegraphics[width=\linewidth]{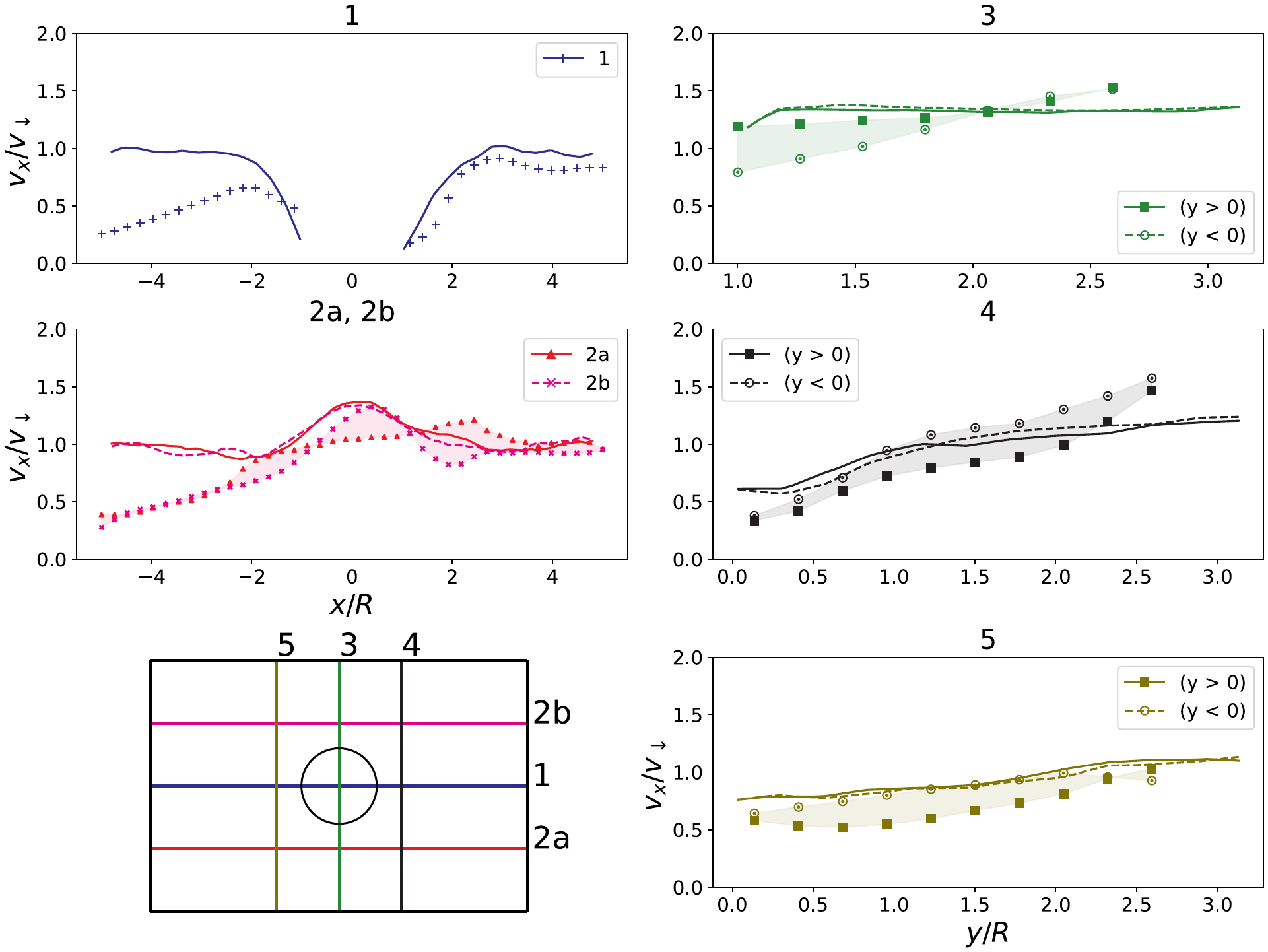}
        \caption{}
        \label{fig:exp_comp_ref_x}
      \end{subfigure}
      \hfill
      \begin{subfigure}[t]{0.49\linewidth}
        \includegraphics[width=\linewidth]{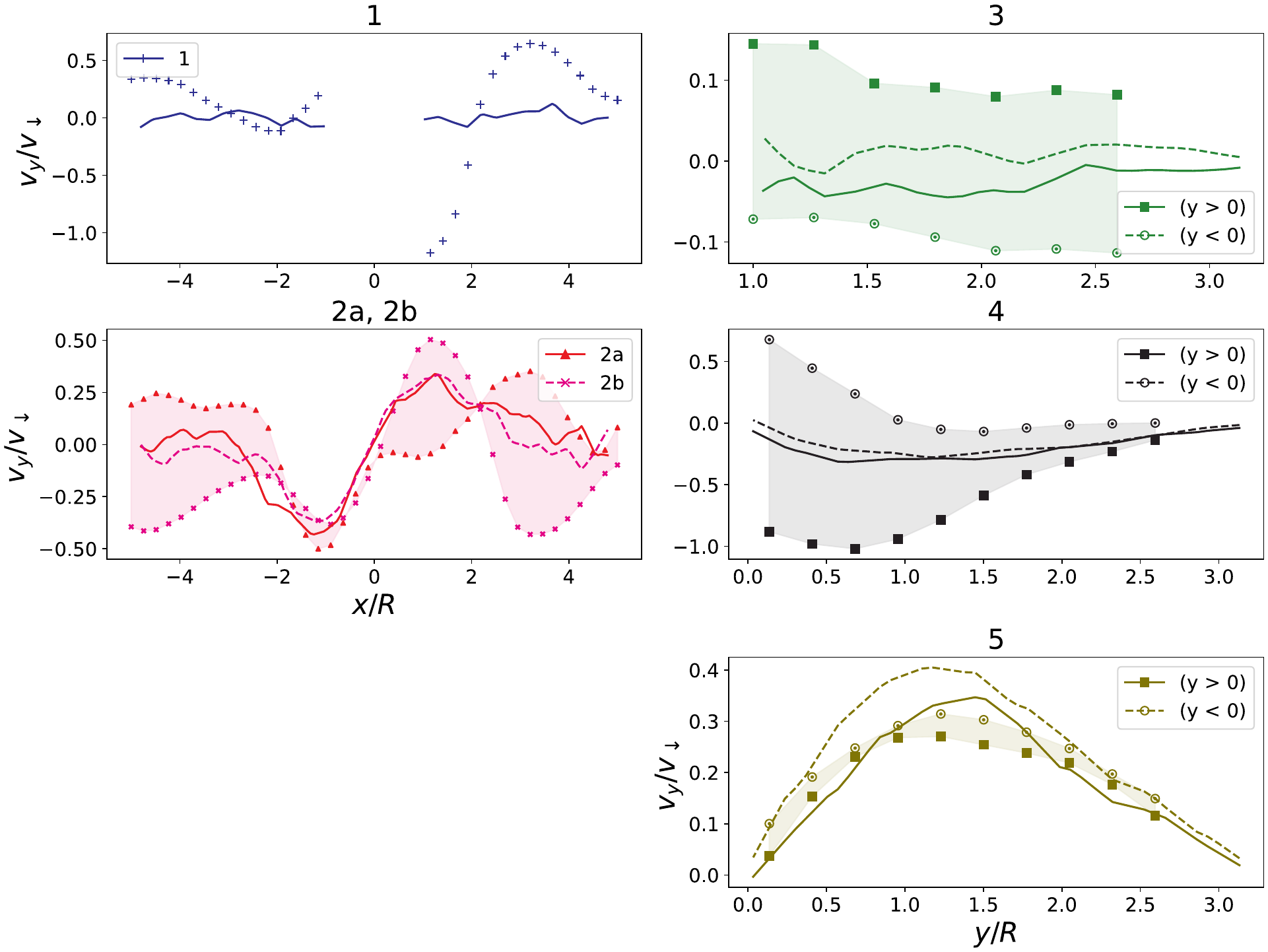}
        \caption{}
        \label{fig:exp_comp_ref_y}
      \end{subfigure}
      \caption{Comparison of velocity profiles between simulation and experiment along the axes illustrated in the graph at the bottom left corner of panel (\subref{fig:exp_comp_ref_x}). Data points with symbols (crosses, squares, circles, etc.) represent experimental measurements, while solid or dashed lines correspond to simulation results. Panel \subref{fig:exp_comp_ref_x} displays the $x$-component of velocity along the axes, and panel \subref{fig:exp_comp_ref_y} shows the $y$-component. All velocities are normalized by the downstream velocity ($v_{\downarrow}$), as defined in the text, and the length along the axis is given in units of the obstacle radius $R$. Overall, there is good agreement between simulation and experiment: in most cases, the simulated curves lie within the range of the experimental data (the shaded region). The main exceptions occur at the beginning of axes $1$, $2a$, and $2b$, which we attribute to experimental limitations in establishing a steady flow. The simulated curves for the vertical axes ($3$, $4$, and $5$) extend slightly further than the experimental curves because the spatial window used to generate the velocity field in the experiment is larger than that used in the simulations. In the simulations, a smaller spatial window could be used due to the larger time average available.}
      \label{fig:exp_comp_ref}
\end{figure*}
\section{Comparison with experiment}
\label{sec:comp_exp}
Lastly, we compared our simulation results with experimental data from an \textit{in vitro} study of Madin–Darby canine kidney epithelial cells migrating through a Stokes geometry, as reported in Durande’s PhD thesis~\cite{duran_2020}, available online at \href{https://theses.fr/2020UNIP7059}{theses.fr}. In the experiment, one half of the channel is initially blocked, while cells are seeded in the remaining half to form a dense, confluent monolayer. The barrier is then removed, allowing the cell sheet to migrate into the previously unoccupied region and initiating the experiment (see \cref{fig:melina}). More specifically, we compared velocity profiles obtained from the temporally averaged, coarse-grained velocity field along linear axes. Given a Cartesian coordinate system with origin at the obstacle center, and letting $R$ be the 
obstacle radius, the axes used are specified in \cref{tab:axis_def} and illustrated at the bottom left corner of \cref{fig:exp_comp_ref_x}. All axes are defined in the region $-5R \leq x \leq 5R,~ -h/2 \leq y \leq h/2$, where $h$ is the channel width. When expressed in units of the typical cell size, the channel dimensions in the simulations and experiments are approximately the same.
\begin{table}[h]
    \centering
    \caption{Definition of axes used for velocity profile comparison. $R$ is the obstacle radius}
    \label{tab:axis_def}
    \begin{tabular*}{0.48\textwidth}{@{\extracolsep{\fill}}lll}
        \hline
        Axis & Equation & Description \\
        \hline
        1   & $y = 0$         & Horizontal axis through center \\
        2a  & $y = -1.66 R$   & Lower horizontal axis \\
        2b  & $y = +1.66 R$    & Upper horizontal axis \\
        3   & $x = 0$         & Vertical axis through center \\
        4   & $x = +1.66 R$    & Right vertical axis \\
        5   & $x = -1.66 R$   & Left vertical axis \\
        \hline
    \end{tabular*}
\end{table}
In order to make the comparison possible, velocity profiles are normalized by the mean velocity value in a region sufficiently far from the obstacle downstream ($v_{\downarrow}$), where the flow is well established. Specifically, $v_{\downarrow}$ is the $x$-component of the average velocity in the region $6.53 R \leq x \leq 6.67 R$. To obtain the profiles, we first generate the velocity field in the same manner as for the output measurements, and then the field is interpolated along the axes. The experimental procedure is similar, because the data extracted from the experiment consist of cell positions over time. 
Aiming to reproduce the experimental results, we executed a simulation with $\tau=1.25\tau_{0}$,  $k_\mathrm{adh}=5.556 \epsilon/\sigma^2$, and $F_i=4.4v_0/\mu$. These parameter values were selected based on the results from the exploration of extreme configurations, with the objective of achieving relatively high alignment and a density close to confluence. The simulation and experimental data are compared in \cref{fig:exp_comp_ref}. Markers represent the experimental results, while solid and dashed curves are simulated results. \Cref{fig:exp_comp_ref_x} shows the $x$-component velocity profiles along the axes, while \cref{fig:exp_comp_ref_y} shows the $y$-component. For each vertical axis, the two halves ($y < 0$ and $y > 0$) are superimposed in the same graph, with one half multiplied by $-1$, because asymmetries can be seen more easily in that way. For the same reason, axes $2a$ and $2b$ are also superimposed in the same graph, with one of them flipped. In the majority of cases, the agreement between the simulations and the experimental data is reasonable, with the computational data almost always falling within the region defined by the two experimental curves. The exceptions are the data for the horizontal axes in the $x<0$ region. We attribute this discrepancy to the experiment’s insufficient duration for establishing steady-state flow in this region. Unlike the simulations, which have a virtually unbounded supply of cells, the experiment contains only a finite number of cells that can flow through the channel.
Furthermore, the experiment shows a significant velocity asymmetry immediately downstream of the obstacle, with cells preferentially moving from top to bottom in this region. This effect can be observed on axis $1$ for the $y$-component velocity (\cref{fig:exp_comp_ref_y}), where the values are strongly negative in the region behind the obstacle ($x \approx 1R$). This behavior was not observed in the stationary state for the simulation in this section, although a similar asymmetry was observed in the simulation for the \case{1}{0}{0}, which was analyzed in \Cref{app:100} of the SI.

\section{Discussion and conclusion}

We have benchmarked the Deformable Active Membrane Model (DAMM) in a confined-flow geometry designed to probe epithelial monolayer migration around a circular obstacle. In the DAMM, cells are represented by deformable adhesive boundaries whose polarity relaxes toward their center-of-mass velocity. This combination of explicit cell interfaces, mechanical deformability, and velocity-polarity feedback enables the model to capture a broad range of collective behaviors within a single computational framework.

Across the explored parameter space, the DAMM generated states ranging from disordered, liquid-like flows to strongly aligned, solid-like motion. Rapid self-alignment is the main mechanism promoting collective order, while adhesion and inlet forcing modulated tissue cohesion, density, and neighbor rearrangements. Compared with the Multiparticle Model (MPM), the DAMM reached substantially higher alignment, exhibited a more systematic increase in density with inlet forcing, and accessed states closer to both limits of the solid--liquid spectrum. These results show that explicit deformable boundaries and self-alignment can produce robust collective organization without requiring direct orientational alignment between neighboring cells.

Our results further identify cell deformability as an important regulator of tissue transport. Increasing the target shape index $p_0$ produced more elongated membrane shapes and enhanced the mean flow velocity through the channel. More deformable cells can accommodate local stresses and modify their shapes while moving through crowded regions, thereby facilitating rearrangements and collective transport. The DAMM therefore provides a direct link between cell-scale mechanical properties, cell morphology, and tissue-scale migration.

The comparison with migrating MDCK epithelial monolayers provided an additional test of the model. For suitable values of the alignment time $\tau$, adhesion strength $k_{\mathrm{adh}}$, and inlet force $F_i$, the simulated velocity profiles qualitatively reproduced the experimental profiles along several horizontal and vertical transects around the obstacle. The principal discrepancies occurred upstream of the obstacle and in the asymmetric transverse motion observed experimentally. These differences may partly arise because  the simulations were analyzed in a statistically stationary regime, whereas the experiment involved a finite cell sheet evolving over a limited observation time. Although asymmetric flows also emerged for some model parameters, their spatial structure differed from that measured experimentally. Determining whether these asymmetries originate from transient dynamics, boundary conditions, fluctuations, or additional biological mechanisms remains an important direction for future work.

The present model deliberately omits several features of real epithelial tissues, including cell division, substrate-dependent friction, intracellular polarity dynamics, and active regulation of cell-cell contacts. Incorporating these ingredients may improve quantitative agreement with experiments and clarify how biochemical and mechanical processes interact to shape migration around obstacles. Further comparisons across obstacle sizes, channel widths, and independently measured tissue parameters would also help determine which aspects of the flow are universal and which depend on system-specific details.

A promising extension of the DAMM would be to incorporate the active regulation of membrane contractility in response to biochemical and mechanical cues. Experiments on neural crest cells have shown that collective chemotaxis can be driven by polarized supracellular actomyosin contraction, with enhanced contractility at the rear of the migrating group promoting cell rearrangements and forward motion~\cite{shellard_supracellular_2018}. Owing to its explicit representation of cell boundaries, the DAMM provides a natural framework in which such mechanisms could be introduced, for instance through spatial and temporal modulation of the tension of membrane bonds at the tissue boundary. The model could further be coupled to an external substrate field to investigate collective durotaxis, including the response to, and remodeling of, local substrate stiffness. Recent experiments indicate that neural crest cells migrate along a self-generated stiffness gradient and that durotactic and chemotactic cues act cooperatively to polarize the actomyosin machinery and guide collective migration~\cite{shellard_collective_2021}. Incorporating these feedback would allow the DAMM to address how cell deformability, active contractility, chemical signaling, and environmental mechanics jointly determine collective migration.

To facilitate such developments, the DAMM is implemented in the open-source package \textit{Phystem}, which combines a Python interface with a high-performance computational core and provides tools for simulation recovery, automated measurements, visualization, and data generation.

Overall, our findings establish self-aligning deformable active membranes as a versatile framework for investigating collective migration in confined and heterogeneous environments. By connecting polarity feedback, adhesion, density, and cell-shape adaptation to emergent tissue flow, the DAMM offers a mechanically explicit route for studying how cellular properties control the large-scale dynamics of active biological materials.

\section*{Data availability}
The data supporting the findings of this study are available within the article and its Supplementary Information. The simulation code used to generate all the results is implemented in the open-source Phystem package and is publicly available at \url{https://github.com/marcos1561/phystem}. The supplementary movies associated with this work are publicly available on Zenodo at \href{https://doi.org/10.5281/zenodo.23084016}{https://doi.org/10.5281/zenodo.23084016}.

\section*{Conflicts of interest}
There are no conflicts of interest to declare

\section*{Acknowledgments}
The authors would like to thank the \href{https://pnipe.mcti.gov.br/search?term=VD%20Lab}{VD Lab cluster} at IF-UFRGS for the computational support with the simulations, the Brazilian agencies CNPq for funding through the International Cooperation Project (process 443517/2023-1), and CAPES for the fellowship (Finance Code 001). We are also grateful to Mélina Durande for kindly granting permission to use data from her PhD thesis.


\newpage


\bibliography{references.bib} 
\bibliographystyle{rsc} 

\end{document}


\section{Coarse-Graining Method}
\label{app:coarse_grainig}
\begin{figure}[h!]
    \centering
    \includegraphics[width=\linewidth]{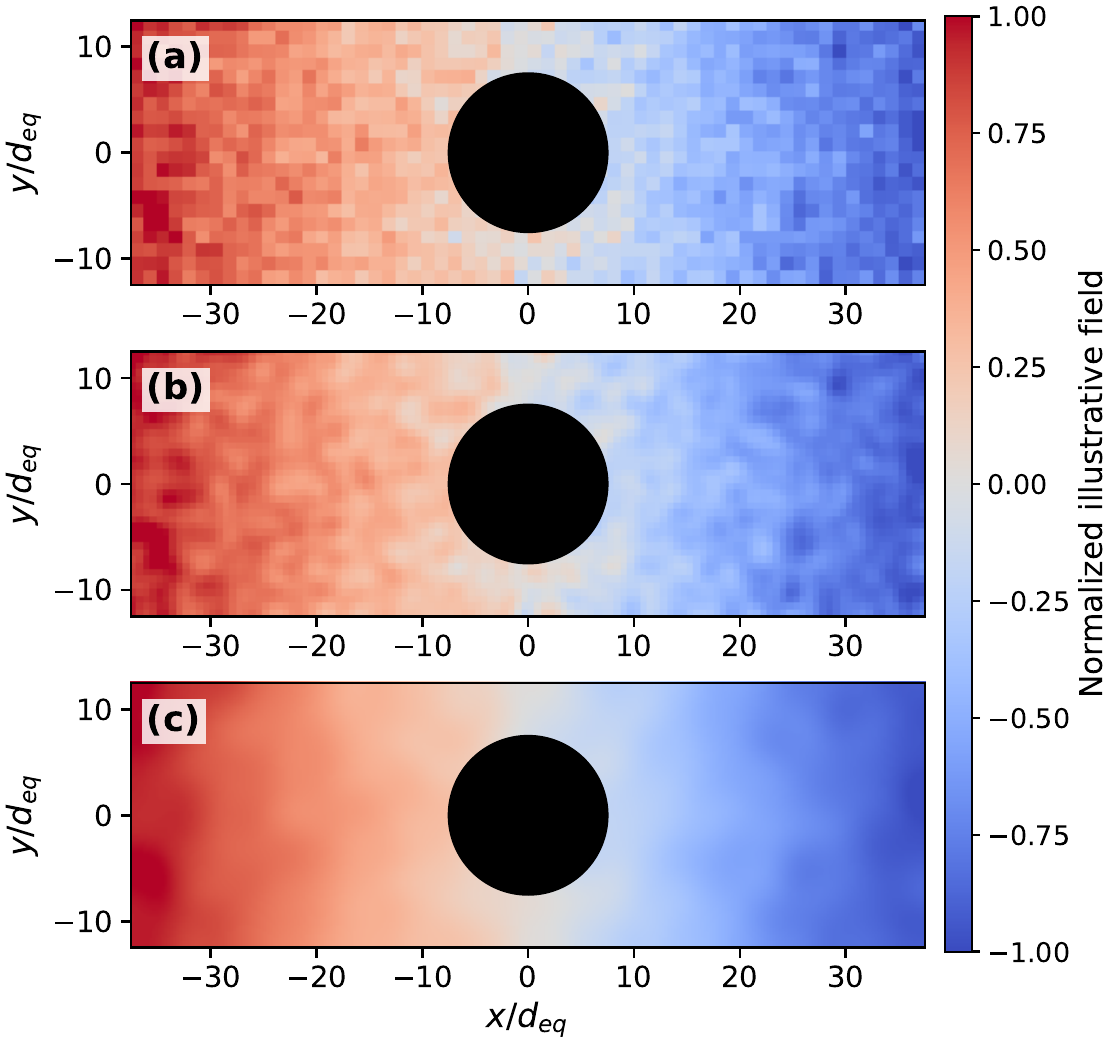}
    \caption{
Illustration of the successive processing steps applied to density fields.
Panel~\textbf{(a)} shows the raw density field represented as a color map,
panel~\textbf{(b)} shows the field after the application of a Gaussian filter,
and panel~\textbf{(c)} shows the filtered field after interpolation onto a
finer grid. In all panels, the black circle represents the channel obstacle, and the color scale indicates the local value of the density field. 
    }
    \label{fig:den_steps}
\end{figure}

In this section we will describe how we compute spatial fields from simulation data. First we divide the space into non-overlapping rectangular windows of side length $2d_{eq}$. For each window, the density is calculated from the number of membrane centers contained within the window, while the velocity field is obtained by averaging the velocities of those membranes. Furthermore, we convolve the obtained density fields with a Gaussian filter in space with $\sigma_G=1.2d_{eq}$, in order to reduce spatial fluctuations and, finally, the data are linearly interpolated to a finer grid of points (windows of side length $0.2 d_{eq}$) than the original grid of points, thus generating a smoother field (see \cref{fig:den_steps}).

\section{Additional Simulations compared to Experimental Data}
\label{app:more_sims}
\begin{figure}[h!]
    \centering
    \includegraphics[width=0.8\linewidth]{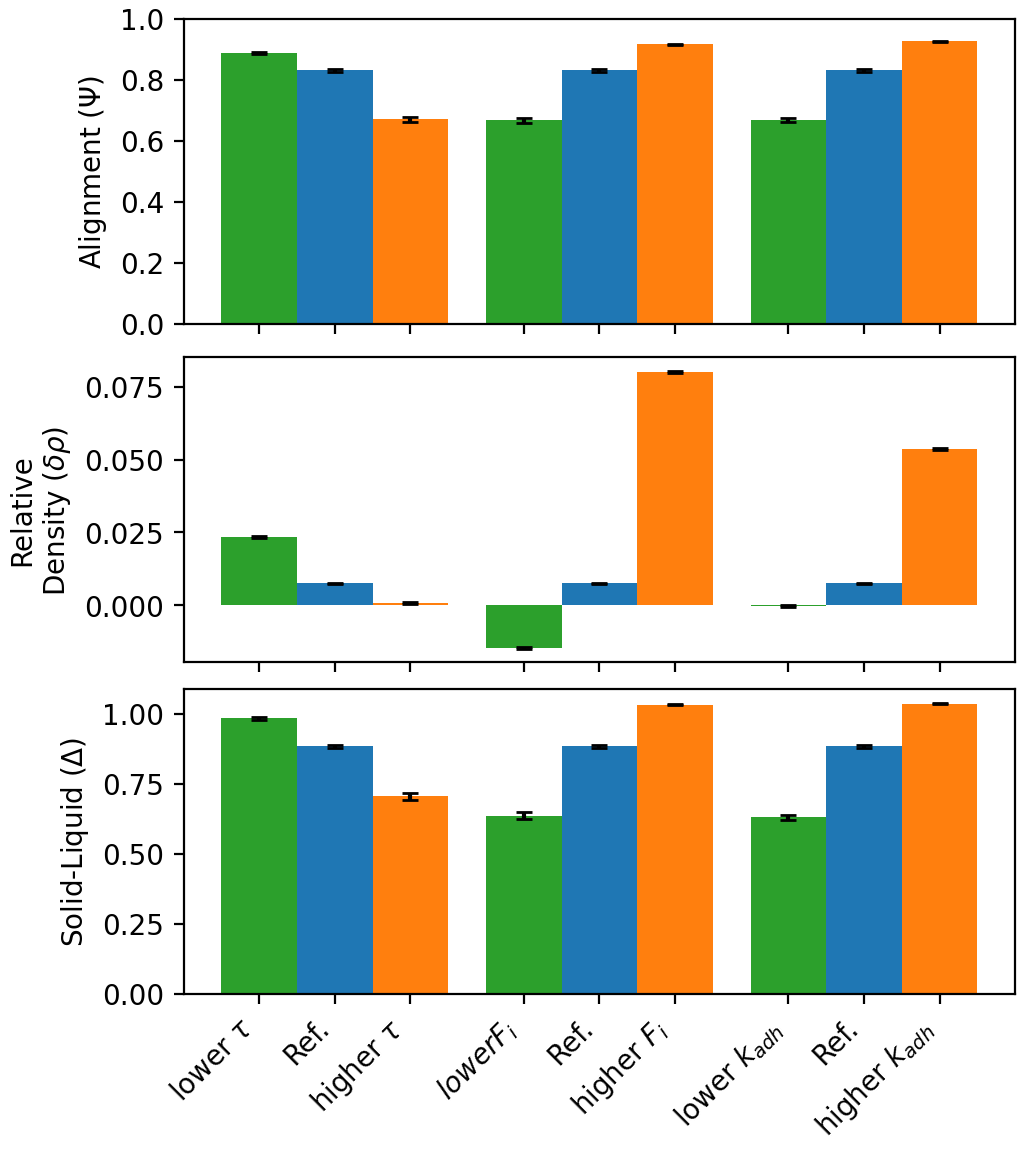}
    \caption{Input measurement values, visualized as vertical bars, for all simulations around the reference. Each graph is related to an input measurement and has three groups, each consisting of three vertical bars. Each group refers to the variation of one parameter ($\tau$, $F_i$ and $k_\mathrm{adh}$, in that order) and each vertical bar represents a simulation with a given set of parameters, in such a way that the varied parameter has a lower value in the leftmost bar (green bar), the reference value in the middle (blue bar) and a higher value in the rightmost bar (orange bar). Error bars represent the standard error of the mean. The parameter values are shown in \cref{tab:exp_sim_par_values}.}
    \label{fig:exp_comp_input_value}
\end{figure}

To verify the robustness and the sensitivity of the results we obtained in \Cref{sec:comp_exp} with respect to model parameters, we executed simulations systematically varying $\tau$, $F_i$ and $k_\mathrm{adh}$ around the values used in the simulation of that section, referred to here as the reference simulation. The chosen parameter values are listed in \cref{tab:exp_sim_par_values}. To verify that the variation actually changed the physical properties of the tissue, input measurements were calculated. \Cref{fig:exp_comp_input_value} shows the results for all input measurements. Each graph has three groups of three bars, with each group associated with one of the explored parameters. Within each group, the green bar (on the left) represents the simulation with the parameter value below the reference value, the blue bar (in the center) corresponds to the reference simulation, and the orange bar (on the right) represents the simulation with the parameter value above the reference. In all cases, we observe a significant and monotonic change in the input measurement associated with the parameter under investigation, indicating that variations in the parameter lead to real and systematic alterations in the system’s physical properties. Moreover, this analysis provides an estimate of the sensitivity of the input measurements to each varied model parameter.

\begin{table}[h]
    \centering
    \caption{Values of the explored parameters for the 6 simulations (rows 2 to 7) performed around a reference simulation (first row) plus the values used in the \case{1}{0}{0} (last row), which is an exceptional case that will be analyzed in \Cref{app:100}. The symbol ``-'' indicates that the value is the same as the reference}
    \label{tab:exp_sim_par_values}
    \begin{tabular*}{0.48\textwidth}{@{\extracolsep{\fill}}llll}
        \hline
        Simulation & $\tau~(\tau_0)$& $k_\mathrm{adh}~(\epsilon / \sigma^2)$  & $F_i~(v_0/\mu)$\\ \hline
        Reference & $1.25$ & $5.556$ & $4.4$  \\ 
        Higher $\tau$ & $1.4$ & - & - \\ 
        Lower $\tau$ & $1.1$ & - & - \\ 
        Higher $k_\mathrm{adh}$ & - & $11.111$ & - \\ 
        Lower $k_\mathrm{adh}$ & - & $1.111$ & - \\ 
        Higher $F_i$ & - & - & $6.5$ \\ 
        Lower $F_i$ & - & - & $3.4$ \\ 
        \casenl{1}{0}{0} & $0.5$ & $0.556$ & $1$ \\ \hline
    \end{tabular*}
\end{table}

\begin{figure*}[h!]
  \makebox[\textwidth][c]{%
    \begin{minipage}{\textwidth}
      \centering
      \begin{subfigure}[t]{0.49\linewidth}
        \includegraphics[width=\linewidth]{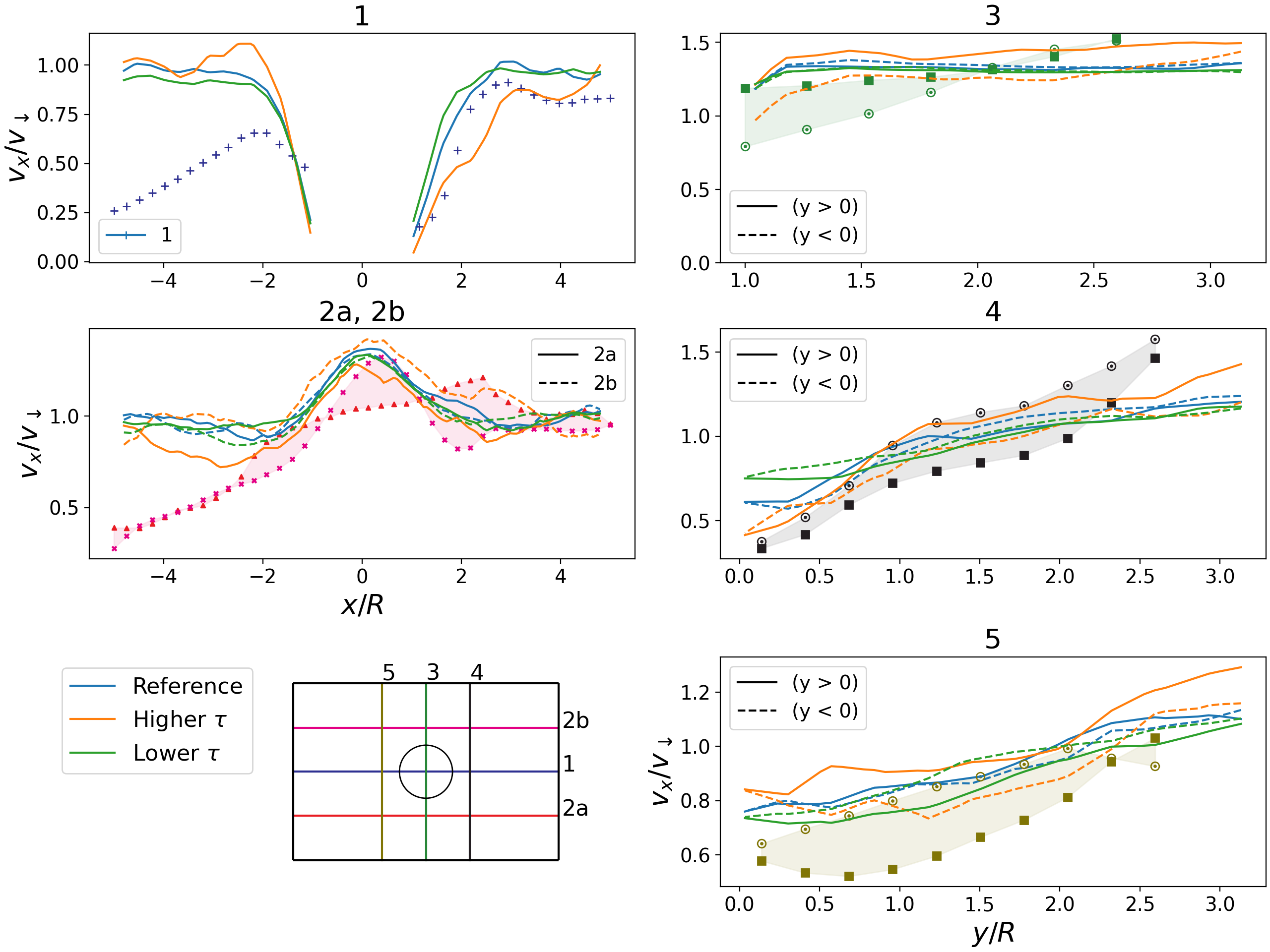}
        \caption{Horizontal velocity component ($v_x$)}
        \label{fig:exp_comp_rt_x}
      \end{subfigure}
      \hfill
      \begin{subfigure}[t]{0.49\linewidth}
        \includegraphics[width=\linewidth]{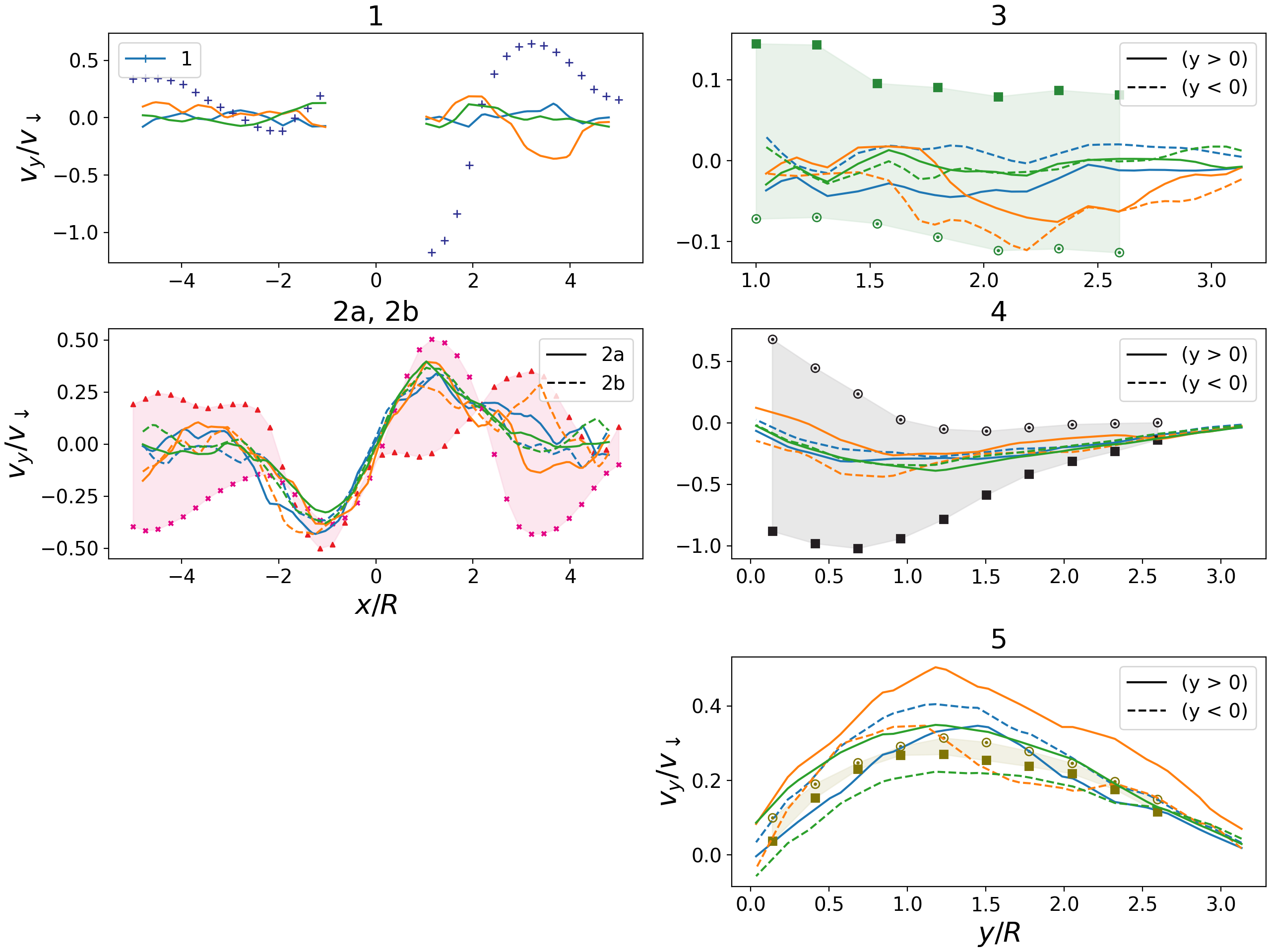}
        \caption{Vertical velocity component ($v_y$)}
        \label{fig:exp_comp_rt_y}
      \end{subfigure}
      \caption{Velocity profiles for the $x$ and $y$ components for simulations in which the relaxation time ($\tau$) was varied. The reference simulation curves (blue curves) and the experimental data (data in symbols) are superimposed with the curves for simulations where the varied parameter value is higher and lower than the reference (orange and green curves, respectively).}
      \label{fig:exp_comp_rt}
    \end{minipage}
  }
\end{figure*}

\begin{figure*}[h!]
  \makebox[\textwidth][c]{%
    \begin{minipage}{\textwidth}
      \centering
      \begin{subfigure}[t]{0.49\linewidth}
        \includegraphics[width=\linewidth]{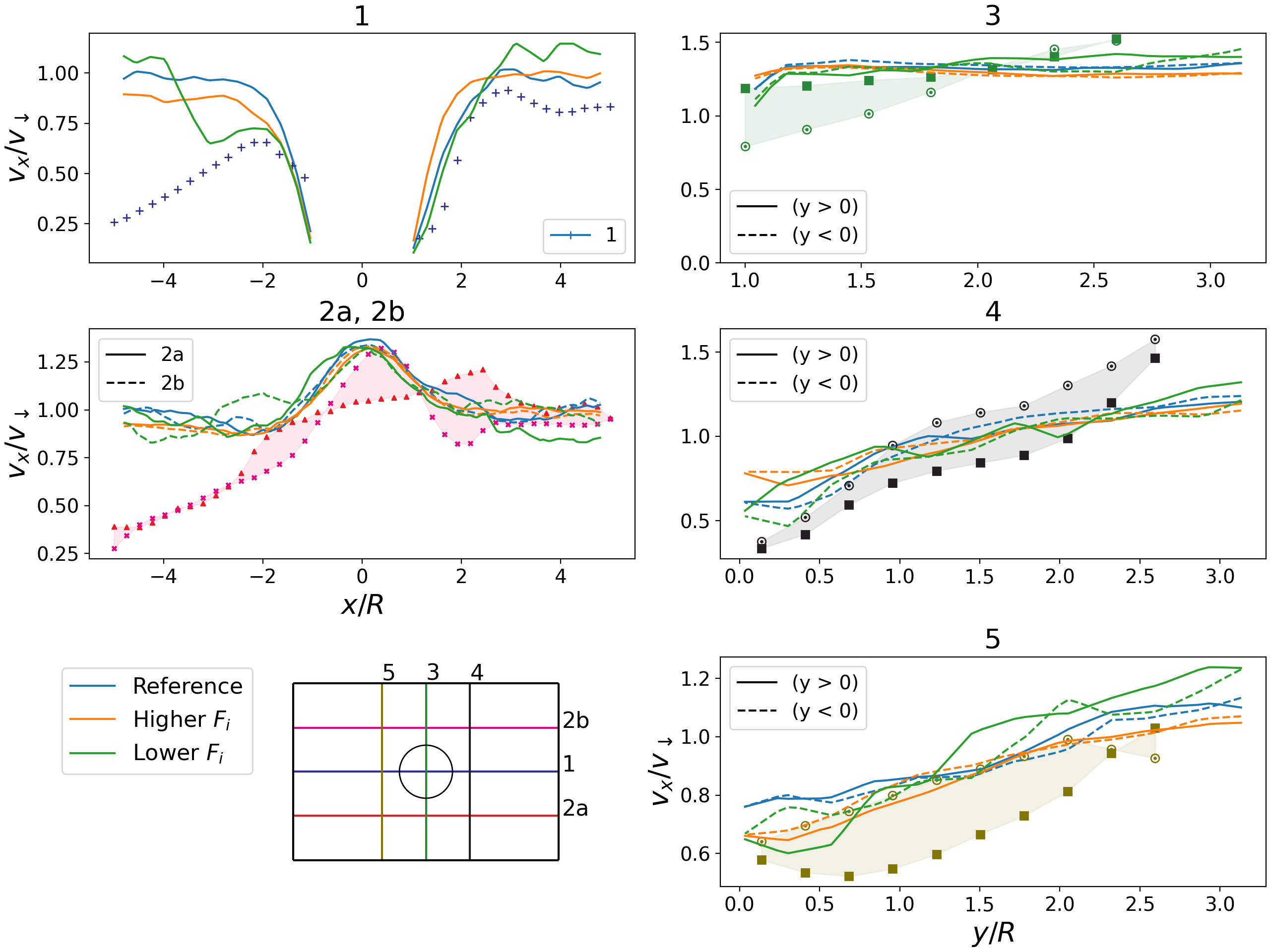}
        \caption{Horizontal velocity component ($v_x$)}
        \label{fig:exp_comp_flux_x}
      \end{subfigure}
      \hfill
      \begin{subfigure}[t]{0.49\linewidth}
        \includegraphics[width=\linewidth]{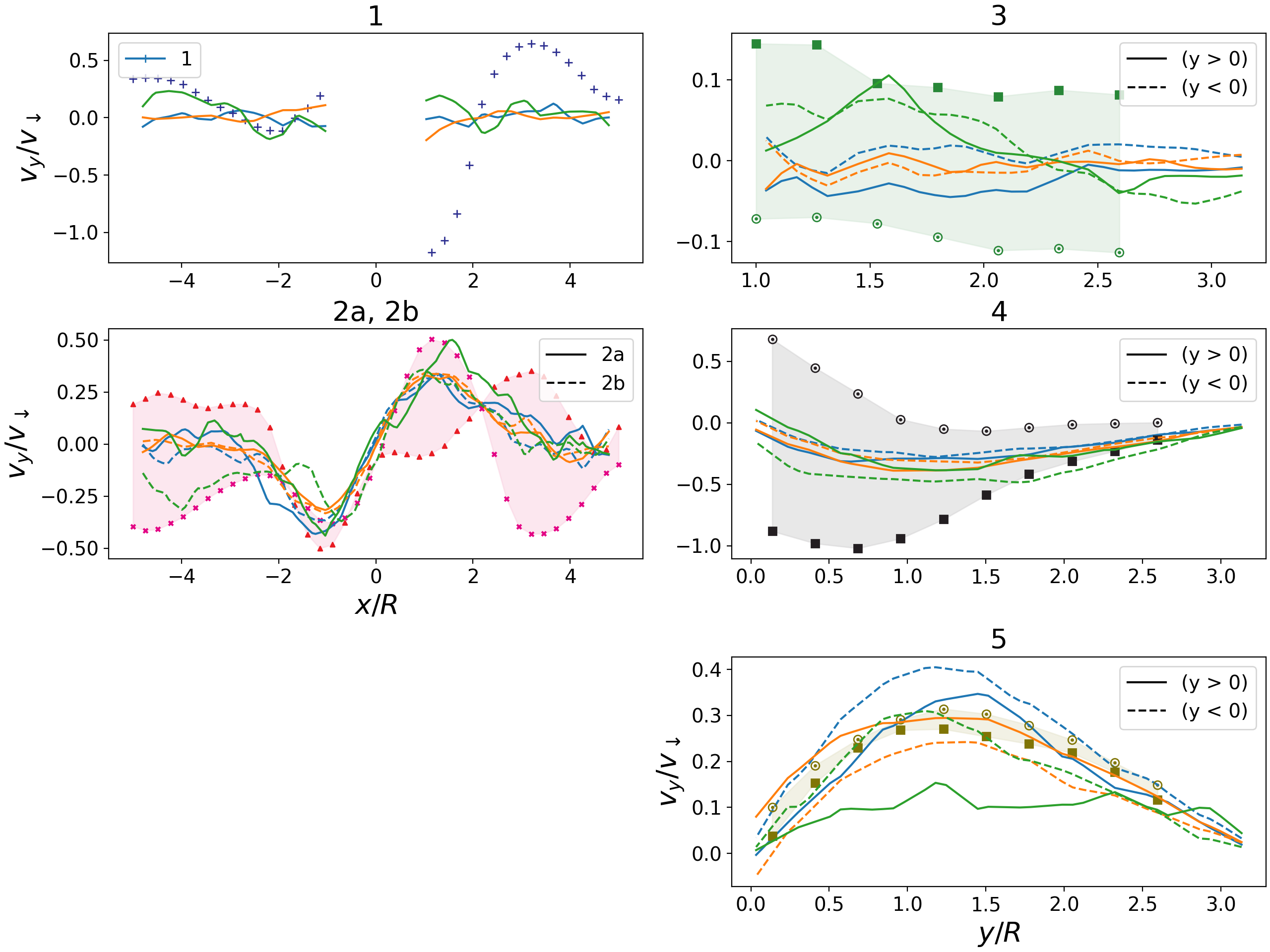}
        \caption{Vertical velocity component ($v_y$)}
        \label{fig:exp_comp_flux_y}
      \end{subfigure}
      \caption{Velocity profiles for the $x$ and $y$ components for simulations in which the flux force ($F_i$) was varied. The reference simulation curves (blue curves) and the experimental data (data in symbols) are superimposed with the curves for simulations where the varied parameter value is higher and lower than the reference (orange and green curves, respectively).}
      \label{fig:exp_comp_flux}
    \end{minipage}
  }
\end{figure*}

\begin{figure*}[h!]
  \makebox[\textwidth][c]{%
    \begin{minipage}{\textwidth}
      \centering
      \begin{subfigure}[t]{0.49\linewidth}
        \includegraphics[width=\linewidth]{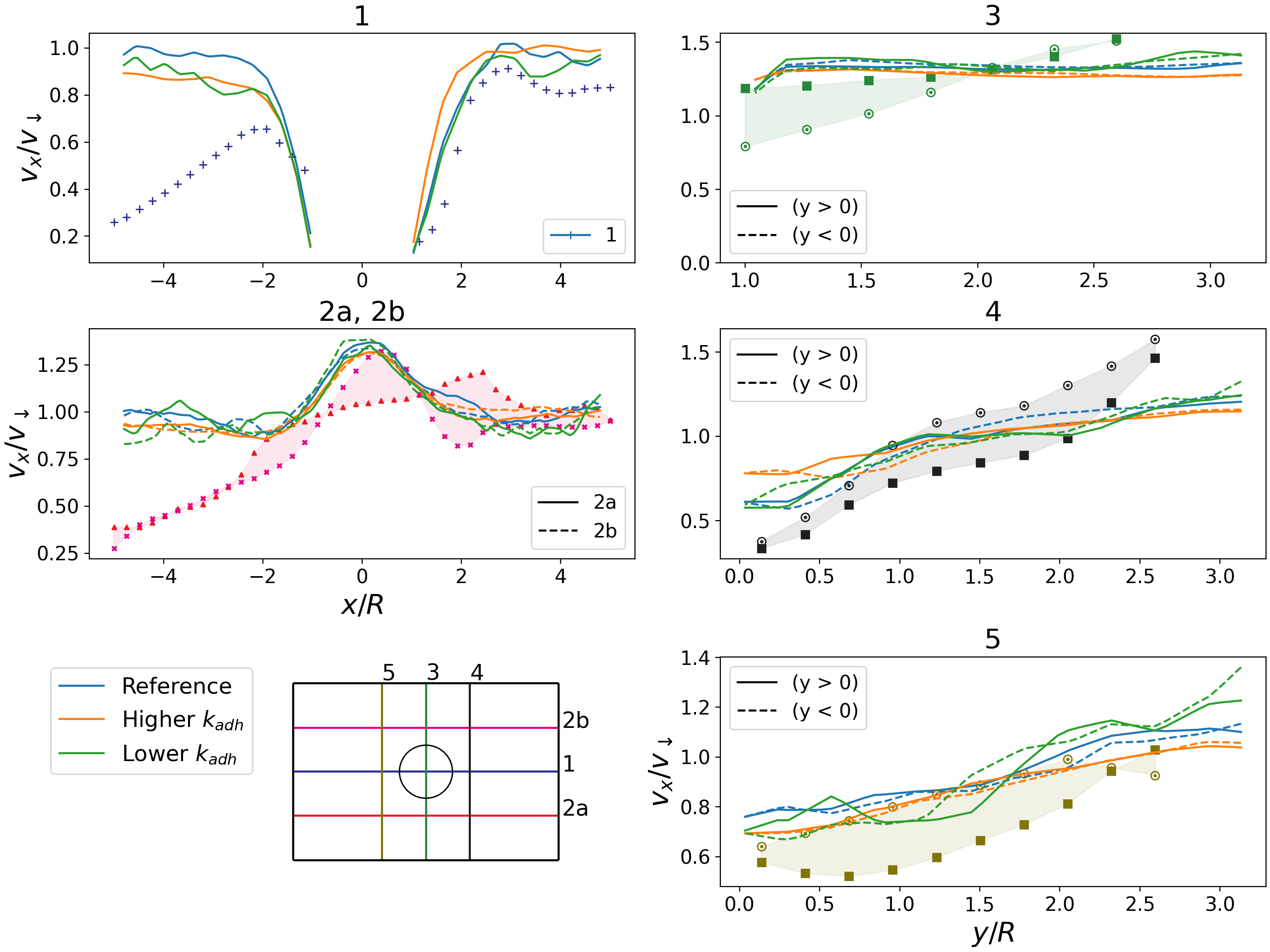}
        \caption{Horizontal velocity component ($v_x$)}
        \label{fig:exp_comp_adh_x}
      \end{subfigure}
      \hfill
      \begin{subfigure}[t]{0.49\linewidth}
        \includegraphics[width=\linewidth]{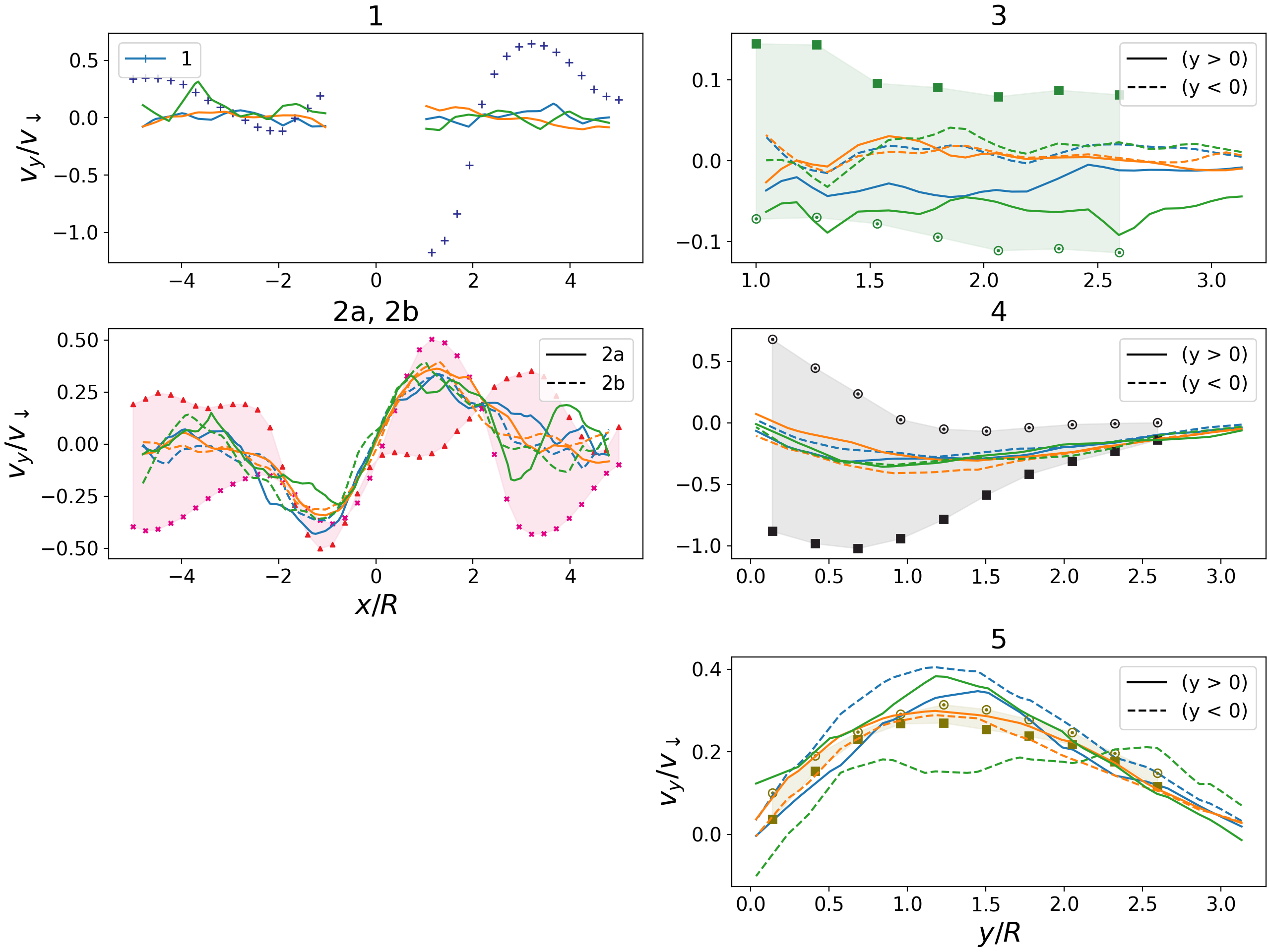}
        \caption{Vertical velocity component ($v_y$)}
        \label{fig:exp_comp_adh_y}
      \end{subfigure}
      \caption{Velocity profiles for the $x$ and $y$ components for simulations in which the adhesion ($k_\mathrm{adh}$) was varied. The reference simulation curves (blue curves) and the experimental data (data in symbols) are superimposed with the curves for simulations where the varied parameter value is higher and lower than the reference (orange and green curves, respectively).}
      \label{fig:exp_comp_adh}
    \end{minipage}
  }
\end{figure*}

\Cref{fig:exp_comp_rt,fig:exp_comp_flux,fig:exp_comp_adh} show the velocity profile comparisons for simulations in which $\tau$, $F_i$, and $k_\mathrm{adh}$ were varied around the reference values. Each figure displays the experimental data alongside the simulated curves for the reference case, as well as for simulations with the corresponding parameter set above and below the reference value, all overlaid on the same graph. In most cases, the simulation results remain consistent with the experimental data. The largest differences between simulations are observed in the graph for axis $5$ of the $y$-component (lower-right graph of panel (b) in each of \cref{fig:exp_comp_rt,fig:exp_comp_flux,fig:exp_comp_adh}), where noticeable variations in the peak values of the curves can be seen. On this axis, the agreement with the experimental data improves for lower $\tau$, higher $F_i$, and higher $k_\mathrm{adh}$ compared to the reference, with the best match observed for higher $k_\mathrm{adh}$ (orange curve in \cref{fig:exp_comp_adh_y}). These cases consistently correspond to simulations with greater alignment than the reference (as shown in \cref{fig:exp_comp_input_value}), indicating that the cellular tissue exhibits a highly aligned flow.

\section{Comparing the \case{1}{0}{0} with Experimental Data}
\label{app:100}
\begin{figure*}[h!]
  \makebox[\textwidth][c]{%
    \begin{minipage}{\textwidth}
      \centering
      \begin{subfigure}[t]{0.49\linewidth}
        \includegraphics[width=\linewidth]{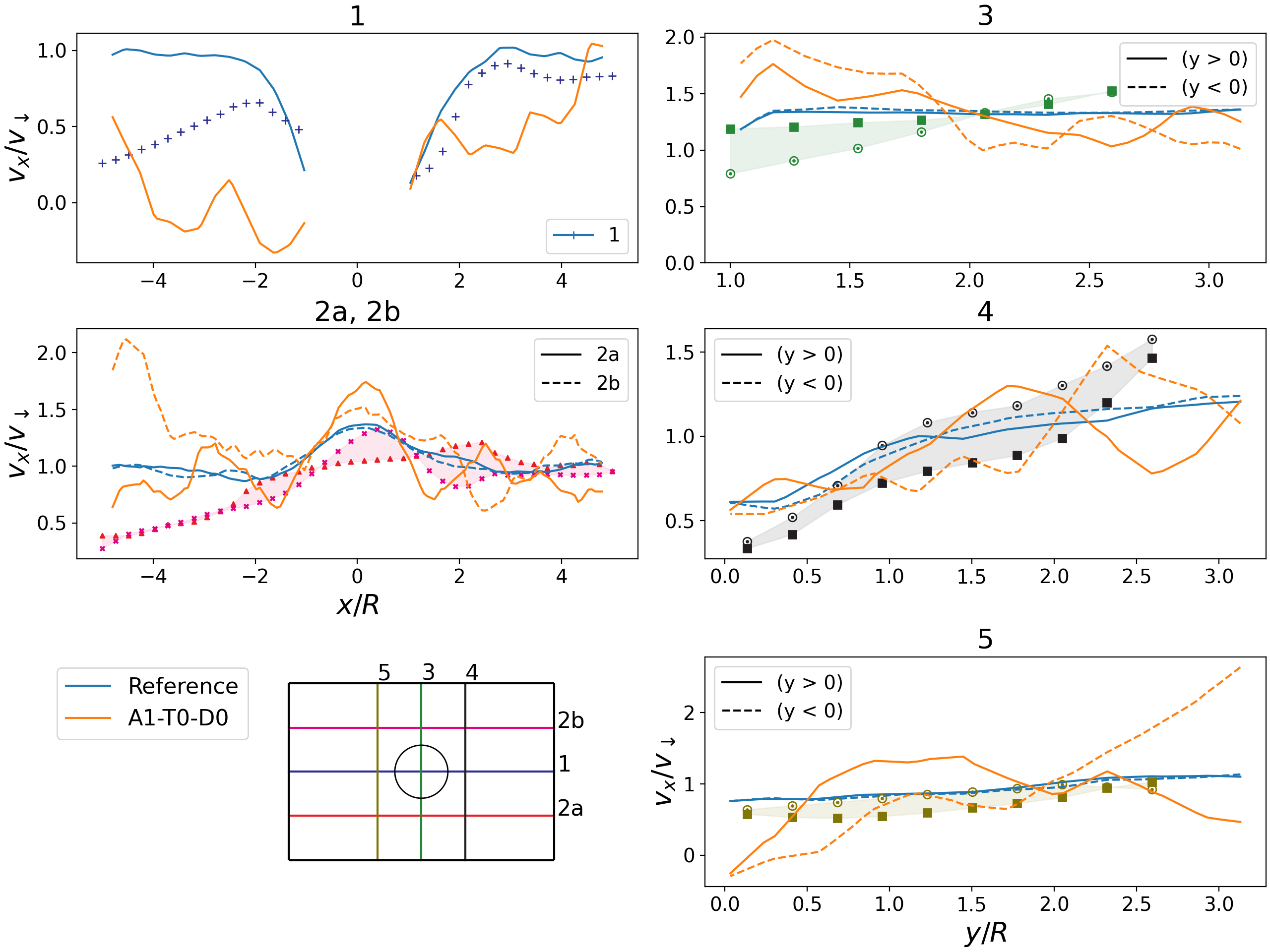}
        \caption{Horizontal velocity component ($v_x$)}
        \label{fig:exp_comp_100_x}
      \end{subfigure}
      \hfill
      \begin{subfigure}[t]{0.49\linewidth}
        \includegraphics[width=\linewidth]{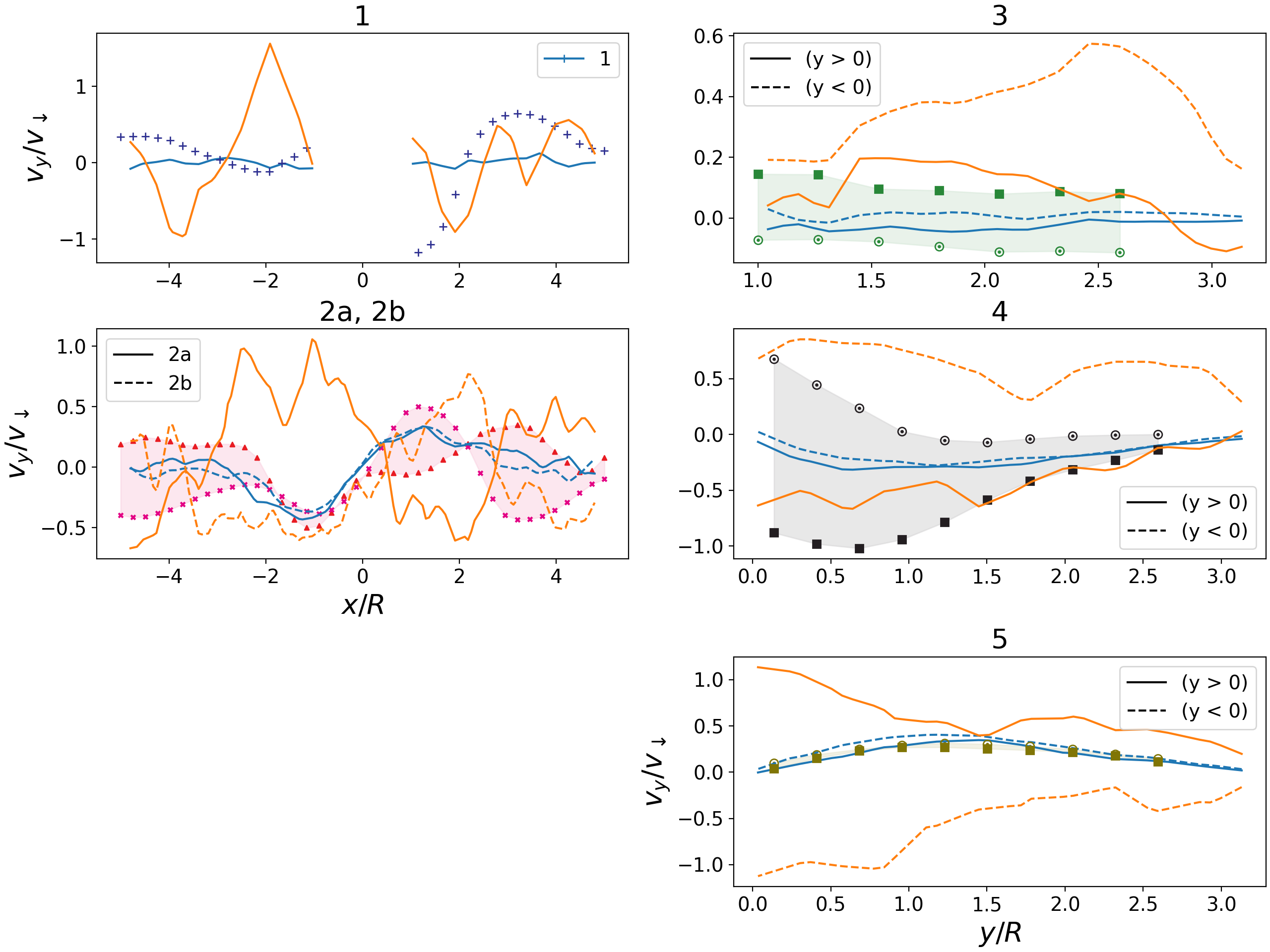}
        \caption{Vertical velocity component ($v_y$)}
        \label{fig:exp_comp_100_y}
      \end{subfigure}
      \caption{Velocity profiles for the $x$ and $y$ components for the \case{1}{0}{0} simulation. The reference simulation curves (blue curves) and the experimental data (data in symbols) are superimposed with the curves for the simulation in question (orange curves).}
      \label{fig:exp_comp_100}
    \end{minipage}
  }
\end{figure*}

Motivated by the asymmetric velocity field of \case{1}{0}{0} (\cref{fig:high_align_output_vel}), we also generated its velocity profiles, as it was the only case in which an asymmetry in the velocity field, similar to that observed in the experiment, was reproduced. The results compared with experimental data and the reference simulation of \Cref{app:more_sims} are in \cref{fig:exp_comp_100}. The simulation curves exhibit more fluctuations than both the experiment and the reference simulation, primarily due to the lower alignment ($\Psi = 0.3$ (\cref{tab:all_input_measurements}) compared to $\Psi \approx 0.8$ in the reference). As a result, the reference simulation provides a closer match to the experimental data, suggesting that the cellular tissue is relatively well ordered under the experimental conditions. The asymmetry noted in the previous discussions can be seen on axis $4$ for $v_y$, at $y/R \approx 0$. Since one of the curves is multiplied by $-1$, the spread between them is related to how asymmetric the field is (in a perfectly symmetric field, both curves should vanish at $y/R=0$). The simulation and the experiment exhibit an asymmetry at that location. For the same reasons, the simulation also has an asymmetry on axis $5$ at $y/R\approx 0$, whereas the experiment does not. These asymmetries can also be observed along axes $2a$ and $2b$ for $v_y$, because an asymmetric field should have curves oscillating in opposite directions (similar to the simulation curves for the \case{1}{0}{0}); however, the experiment only has this behavior after the obstacle ($x/R > 0$).

\section{Anti-Intersection Method}
\label{sec:anti_intersection}
As mentioned in \cref{sec:model}, a potential limitation of the DAMM is the possibility of unphysical membrane overlaps. When such an intersection occurs, the overlapping particles do not naturally feel a mutual repulsive force. To mitigate this issue, we implemented an algorithm that detects intersections and applies a corrective force to all intersecting particles until the overlap is resolved.

An intersection between membranes $j$ and $k$ is defined to exist if the position of any particle belonging to membrane $j$ lies within the polygon formed by the particle centers of membrane $k$, or vice versa. The corrective force applied to an intersecting particle $i$ of membrane $j$ has a constant magnitude $F_\text{inv}$ and acts antiparallel to the area-minimizing force with a sign correction $\mathbf{F}^\text{out}_{i, j} = (-\nabla_{i, j} E_A) \operatorname{sgn}(A_0 - A_j)$ [\cref{eq:energy_a}], where $\operatorname{sgn}$ is the sign function. This directional choice ensures efficiency, as $\mathbf{F}^\text{out}_{i, j}$ is naturally perpendicular to the segment connecting the particle's membrane neighbors ($\mathbf{r}_{i+1, j} - \mathbf{r}_{i-1, j}$) and points outward from the membrane's interior. We used $F_\text{inv} = 56~v_0/\mu$, which is the maximum repulsive force between particles, as defined in \cref{eq:energy_c_adh}.

To evaluate the frequency of these events, we monitored intersections at the end of all simulations performed in this work by simulating the systems for an additional $50~\tau_0$ and recording the number of active overlaps over time. The results demonstrate that intersections were rare; the only configurations yielding a non-zero overlap count were \case{1}{1}{0} and \case{1}{1}{1}, which exhibited a low mean number of active intersections of four and eight, respectively. Note that these are precisely the two cases with both high alignment and high density, conditions under which the formation of a highly dense region is expected.

\section{Phystem: A software library to explore the Deformable Active Membrane Model}
\label{app:phystem}
All simulation results and images showcasing the active membranes were generated using a Python library called Phystem, developed by the authors. Phystem is an open-source software package with a permissive license; its code is available on \href{https://github.com/marcos1561/phystem}{GitHub}, with a documentation page available \href{https://marcos1561.github.io/phystem/docs/introduction.html}{here}.

Phystem was designed to explore arbitrary agent-based dynamical systems, but it has a built-in implementation of the Deformable Active Membrane Model with a Python interface and a C++ core for computationally intensive tasks. We highly recommend that interested readers consult the \href{https://marcos1561.github.io/phystem/docs/systems/ring.html}{documentation page for the DAMM}, as here we will only briefly demonstrate the basic capabilities of Phystem.

\begin{figure}[h]
    \centering
    \includegraphics[width=\linewidth]{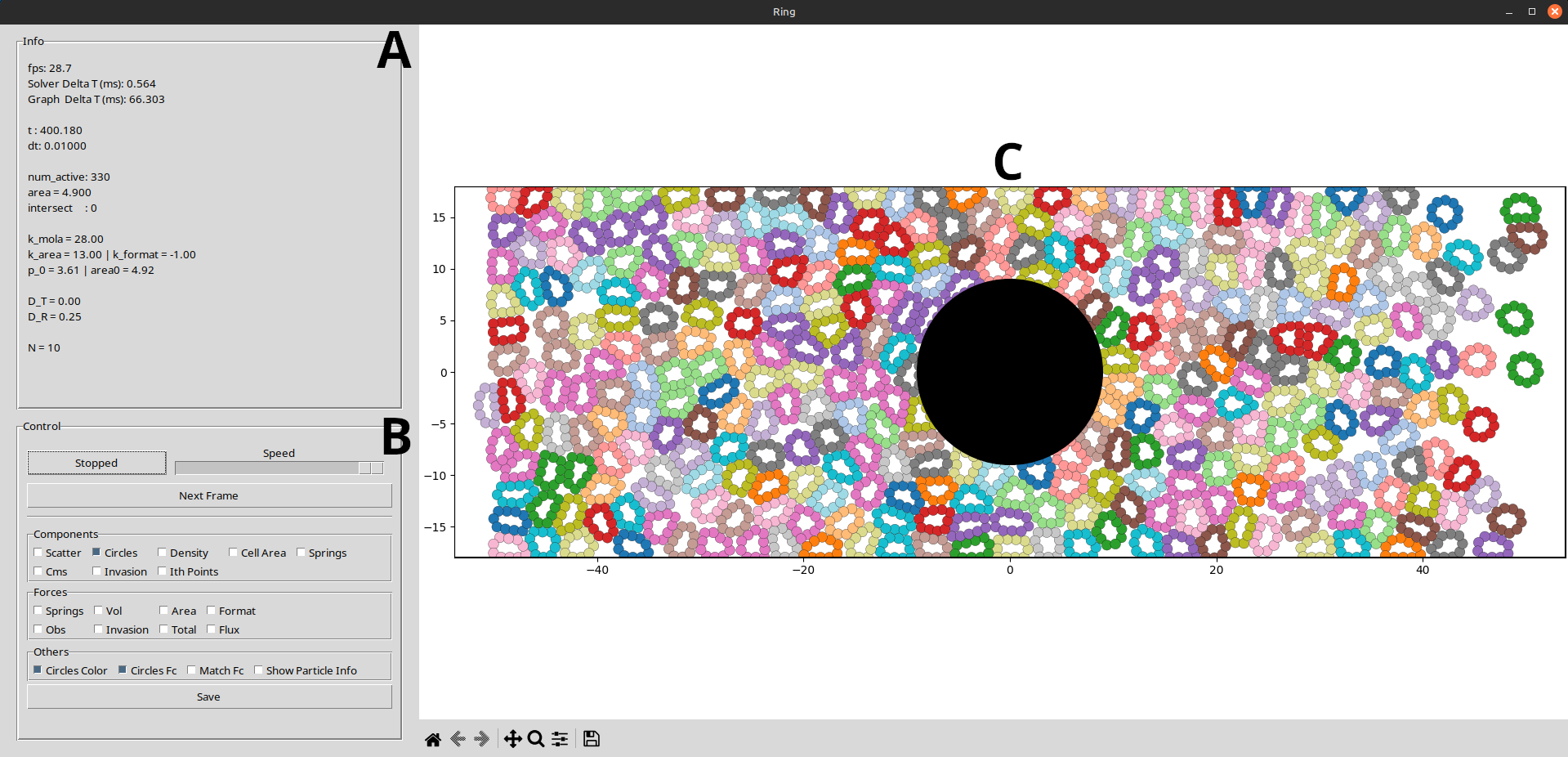}
    \caption{Illustration of the window opened when running Phystem in real-time simulation mode. The interface consists of three main components: Panel A displays key information about the system, such as the elapsed time and the real time per simulation step. Panel B provides controls to pause or resume the simulation, adjust the simulation speed (i.e., the number of simulation steps per frame), toggle visualization aids (such as forces as arrows), and save the current state of the system. Panel C presents the trajectories of the particles as the system evolves over time; this panel is interactive, supporting both zooming and panning.}
    \label{fig:phystem_gui}
\end{figure}

Phystem supports multiple operating modes, the main ones being real-time simulation rendering (M1), video generation (M2), and data collection (M3). After configuring the simulation, mode M1 allows users to observe particle trajectories in real time through an interactive interface (\cref{fig:phystem_gui}). This interface enables pausing or resuming the simulation, displaying system properties, and customizing the visualization, for example, showing the different forces acting on each particle, as illustrated in \cref{fig:model_scheme}. The particle display panel is also interactive, allowing users to zoom, pan, and click on particles to view additional information. Mode M2 can be used to generate a video of the simulation instead of providing real-time visualization.

Mode M3 does not render the simulation; instead, it executes the simulation in the background and collects data. Phystem features a system of collectors, which are objects that define all procedures for collecting specific quantities, ensuring that system integration is completely separated from data collection. This design provides users with greater flexibility and control.

Finally, Phystem includes a serialization/deserialization system, allowing users to save the state of a system and load it later or share it with others. This system also applies to collectors and is used to create checkpoints during mode M3. In the event of an unexpected interruption, users can load a checkpoint and resume data collection without losing all previous progress.

To improve code performance, computationally intensive tasks are executed in a compiled C++ module, and loops are parallelized across CPU cores when possible. The simulations performed in this work proved to be feasible within reasonable execution times. It was possible to simulate up to $10^4$ cells (equivalent to $10^5$ particles, as each cell has $10$ particles) for approximately $10^7$ integration steps, corresponding to execution times on the order of one week on an AMD FX(tm)-8350 Eight-Core Processor, which was sufficient for the flow to stabilize along the channel, even in the disordered regime. In the case of the ordered regime, the time required for flow stabilization was significantly shorter, around $10^5$ steps, corresponding to execution times on the order of hours.

\FloatBarrier
\section{Supplementary Movies}
We created two movies (A and B) for every combination of the selected parameters extremes (\cref{tab:parameter_summary}), showcasing the membranes flowing in the channel. Movie A shows the full simulation channel, whereas movie B provides a magnified view of the region around the circular obstacle. Membranes enter the channel from the left and migrate around the obstacle. Colors indicate asphericity, a scalar measure of shape anisotropy ranging from more circular membranes (dark purple) to more elongated membranes (yellow). The supplementary movies are publicly available on Zenodo at \href{https://doi.org/10.5281/zenodo.23084016}{https://doi.org/10.5281/zenodo.23084016}.

\medskip

\suppmoviepair{mov:A0T0D0}{
Case A0-T0-D0: $\tau=25\,\tau_0$, $k_\mathrm{adh}=0.556\,\epsilon/\sigma^2$, and $F_i=1\,v_0/\mu$.
}

\medskip

\suppmoviepair{mov:A0T0D1}{
Case A0-T0-D1: $\tau=25\,\tau_0$, $k_\mathrm{adh}=0.556\,\epsilon/\sigma^2$, and $F_i=10\,v_0/\mu$.
}

\medskip

\suppmoviepair{mov:A0T1D0}{
Case A0-T1-D0: $\tau=25\,\tau_0$, $k_\mathrm{adh}=33.333\,\epsilon/\sigma^2$, and $F_i=1\,v_0/\mu$.
}

\medskip

\suppmoviepair{mov:A0T1D1}{
Case A0-T1-D1: $\tau=25\,\tau_0$, $k_\mathrm{adh}=33.333\,\epsilon/\sigma^2$, and $F_i=10\,v_0/\mu$.
}

\medskip

\suppmoviepair{mov:A1T0D0}{
Case A1-T0-D0: $\tau=0.5\,\tau_0$, $k_\mathrm{adh}=0.556\,\epsilon/\sigma^2$, and $F_i=1\,v_0/\mu$.
}

\medskip

\suppmoviepair{mov:A1T0D1}{
Case A1-T0-D1: $\tau=0.5\,\tau_0$, $k_\mathrm{adh}=0.556\,\epsilon/\sigma^2$, and $F_i=10\,v_0/\mu$.
}

\medskip

\suppmoviepair{mov:A1T1D0}{
Case A1-T1-D0: $\tau=0.5\,\tau_0$, $k_\mathrm{adh}=33.333\,\epsilon/\sigma^2$, and $F_i=1\,v_0/\mu$.
}

\medskip

\suppmoviepair{mov:A1T1D1}{
Case A1-T1-D1: $\tau=0.5\,\tau_0$, $k_\mathrm{adh}=33.333\,\epsilon/\sigma^2$, and $F_i=10\,v_0/\mu$.
}